\documentclass[conference]{IEEEtran}
\IEEEoverridecommandlockouts
\usepackage{cite}
\usepackage{tabularx}
\usepackage{amsmath,amssymb,amsfonts}
\usepackage{algorithmic}
\usepackage{graphicx}
\usepackage{textcomp}
\usepackage{xcolor}
\usepackage{multirow}
\usepackage{subcaption}
\usepackage{stackengine}
\usepackage{enumitem}
\usepackage{booktabs} 
\newcommand\xrowht[2][0]{\addstackgap[.5\dimexpr#2\relax]{\vphantom{#1}}}

\usepackage{amsmath}
\usepackage{url}
\usepackage{amssymb}
\usepackage{mathtools}
\usepackage{amsthm}
\theoremstyle{plain}
\newtheorem{theorem}{Theorem}[section]

\theoremstyle{definition}
\newtheorem{definition}[theorem]{Definition}

\theoremstyle{remark}

\usepackage[draft,colorinlistoftodos,color=orange!30]{todonotes}

\def\BibTeX{{\rm B\kern-.05em{\sc i\kern-.025em b}\kern-.08em
    T\kern-.1667em\lower.7ex\hbox{E}\kern-.125emX}}
\begin{document}

\title{Neural Causal Graph Collaborative Filtering
}

\author{\IEEEauthorblockN{1\textsuperscript{st} Xiangmeng Wang}
\IEEEauthorblockA{\textit{School of Computer Science} \\
\textit{University of Technology Sydney}\\
Sydney, Australia \\
xiangmeng.wang@student.uts.edu.au}
\and
\IEEEauthorblockN{2\textsuperscript{nd} Qian Li \IEEEauthorrefmark{1}\thanks{\IEEEauthorrefmark{1}: Contributing equally with the first author}\IEEEauthorrefmark{2}}
\IEEEauthorblockA{\textit{School of Elec Eng, Comp and Math Sci} \\
\textit{Curtin University}\\
Perth, Australia \\
qli@curtin.edu.au}
\and
\IEEEauthorblockN{3\textsuperscript{rd} Dianer Yu}
\IEEEauthorblockA{\textit{School of Computer Science} \\
\textit{University of Technology Sydney}\\
Sydney, Australia \\
Dianer.Yu-1@student.uts.edu.au}
\and
\IEEEauthorblockN{4\textsuperscript{th} Wei Huang}
\IEEEauthorblockA{\textit{Deep Learning Theory Team}\\
\textit{RIKEN Center for AIP}\\
Tokyo, Japan\\
Wei.Huang.vr@riken.jp }
\and
\IEEEauthorblockN{5\textsuperscript{th} Qing Li}
\IEEEauthorblockA{\textit{Department of Computing} \\
\textit{Hong Kong Polytechnic University}\\
Hong Kong SAR, China \\
qing-prof.li@polyu.edu.hk}
\and
\IEEEauthorblockN{6\textsuperscript{th} Guandong Xu\IEEEauthorrefmark{2}\thanks{\IEEEauthorrefmark{2}: Corresponding author}}
\IEEEauthorblockA{\textit{School of Computer Science} \\
\textit{University of Technology Sydney}\\
Sydney, Australia \\
guandong.xu@uts.edu.au}
}

\maketitle



\begin{abstract} 
Graph collaborative filtering (GCF) has gained considerable attention in recommendation systems by leveraging graph learning techniques to enhance collaborative filtering (CF). One classical approach in GCF is to learn user and item embeddings with Graph Convolutional Network (GCN) and utilize these embeddings for CF models. 
However, existing GCN-based methods are insufficient in generating satisfactory embeddings for CF models. 
This is because they fail to model complex node dependencies and variable relation dependencies from a given graph, making the learned embeddings fragile to uncover the root causes of user interests.
In this work, we propose to integrate causal modeling with the learning process of GCN-based GCF models, leveraging causality-aware graph embeddings to capture complex causal relations in recommendations.
We complete the task by 1) Causal Graph conceptualization, 2) Neural Causal Model parameterization and 3) Variational inference for Neural Causal Model.
Our Neural Causal Model, called \emph{Neural Causal Graph Collaborative Filtering (NCGCF)}, enables causal modeling for GCN-based GCF to facilitate accurate recommendations.
Extensive experiments show that NCGCF provides precise recommendations that align with user preferences. We release our code and processed datasets at~\url{https://github.com/Chrystalii/CNGCF}.
\end{abstract}


\section{Introduction}
Collaborative Filtering (CF)~\cite{schafer2007collaborative} as an effective remedy has dominated recommendation research for years.
A recent emerging CF paradigm built on graph learning~\cite{xia2021graph}, i.e., Graph Collaborative Filtering (GCF), has been studied extensively~\cite{wang2021graph}.
GCF enhances traditional CF methods~\cite{chen2018matrix,he2017neural} by modeling complex user-item interactions in a graph as well as auxiliary information, e.g., user and item attributes. 
Thus, GCF has shown great potential in deriving knowledge (e.g., user behavior patterns) embedded in graphs.

Generally, GCF models utilize graph representation learning techniques, as described in~\cite{hamilton2020graph}, to derive useful information for downstream CF. These models use graph neural networks to analyze connections and create embeddings, thereby improving CF model optimization.
Graph Convolutional Network (GCN)-based GCF methods leverage GCN's ability to learn local and global information from large-scale graphs, as evidenced by several studies~\cite{wang2019neural,sun2021hgcf,berg2017graph,he2020lightgcn}. 
These methods first acquire vectorized user and item embeddings using a GCN, and then use these embeddings to optimize a CF model, capitalizing on GCN's demonstrated competitive performance in this domain.
For instance, NGCF~\cite{wang2019neural} exploits a GCN to propagate neighboring node messages in the interaction graph to obtain user and item embeddings.
The learned embeddings capture user collaborative behavior and are used to predict preference scores for CF optimization. 
HGCF~\cite{sun2021hgcf} combines GCN with hyperbolic learning to learn embeddings in the hyperbolic space. 
Benefiting from the exponential neighborhood growth in the hyperbolic space, HGCF facilitates learning higher-order user and item relations from the interaction graph.

However, two fundamental drawbacks hinder GCN-based methods from producing satisfactory embeddings.
Firstly, \emph{they ignore distinguishable node dependencies between neighboring nodes and the target node.}
Most GCN-based methods treat all messages from the neighborhood equally, following node commonality~\cite{yan2022two}, which inevitably overlooks the varying dependencies of neighboring nodes to the target node.
However, a user node might have different relations with other neighboring nodes (e.g., item brands), i.e., distinct user preference, which is the essence of personalized recommendations~\cite{10.1145/3485447.3512072}.
As a result, user and item embeddings eventually lose expressive power in the recommendation task, i.e., we cannot know which node is the root cause of user interests.
Secondly, \emph{they lack an explicit encoding of complex relations between variables in the recommendation.} 
Most GCN-based methods assume the co-occurrence of users and items is independent~\cite{xu2023causal}.
However, user preferences are influenced by various variables in real-world recommendations, such as the user conformity caused by user social networks~\cite{zheng2021disentangling}.
Discarding these relations leads to the learned embeddings unable to capture such structural complexity.



Causal modeling sheds light on solving the above drawbacks.
On the one hand, causal modeling identifies intrinsic cause-effect relations between nodes and true user preferences~\cite{wang2022causaldisen}.
For example, we might treat each neighboring node as the cause (e.g., an item brand) and the user preference as the effect in a Causal Graph~\cite{bareinboim2022pearl}.
By estimating the causal effect, we could encode the crucial node dependencies into user and item embeddings to uncover the root causes of user interests.
On the other hand, the Causal Graph is able to model genuine causal relations among the variables in GCFs, capturing variable dependencies inherent in the GCF-based methods.
Those causal relations represent the underlying mechanisms driving the recommendation and can be utilized to guide graph learning toward complex user behaviors.

Given the compelling nature of casual modeling in GCN-based GCFs, in this paper, we aim to integrate GCNs and causal models to facilitate a causality-aware GCF learning.
Motivated by Neural-Causal Connection~\cite{xia2021causal}, this paper proposes to connect GCN learning with the Structural Causal Model (SCM)~\cite{pearl2000models}.
Since the SCM is induced from a Causal Graph and the GCN works on graph-structured data, the integration of the two models becomes practical.
In particular, we first conceptualize the Causal Graph for the SCM, which is built by revisiting existing CFs and padding their limitations in user preference modeling.
Then, we formulate the SCM into a Neural Causal Model, called \emph{Neural Causal Graph Collaborative Filtering (NCGCF)}.
Our NCGCF uses variational inference to approximate structural equations as trainable neural networks, making the learned graph embeddings equally expressive as the causal effects modeled by the SCM.
The integration of causal modeling and graph representation learning offers a novel perspective to facilitate accurate recommendations.
The contributions of this work are:
\begin{itemize}[leftmargin=*]
    \item We complete the Neural-Causal Connection for causal modeling of graph convolutional network in recommendations. 
    \item Our proposed NCGCF is the first Neural Causal Model for graph collaborative filtering, which generates causality-aware graph embeddings for enhanced recommendations.
    \item We validate the effectiveness of our proposed framework through extensive experiments. Our experimental results demonstrate that our approach outperforms existing methods in achieving satisfactory recommendation performance. 
\end{itemize}

\section{Notations and Formulation}
We provide our motivations for defining our Causal Graph.
We give notations that we use throughout the paper.
We give our task formulation, which covers detailed steps toward connecting GCN with the Structural Causal Model. 

 \begin{table*}[]
    \centering
    \caption{Key notations and descriptions.}
    \begin{tabular}{l l}
    \hline
         Notation & Description \\
         \hline \xrowht[()]{5pt}
          $G$ & A Causal Graph\\
         $\mathcal{M}$ & A Structural Causal Model\\
         $\mathcal{M}(\theta)$ & A Neural Causal Model\\
         $\mathcal{V} =\{U,V,E,Y\}$ & Endogenous variables in $G$\\ 
         $\mathcal{F} =\{f_U, f_V, f_E, f_Y\}$ & Structural equations for $G$\\
         $U$, $f_U$ & User variable and its structural equation\\
         $V$, $f_V$ & Item variable and its structural equation\\
        $E$, $f_E$ & Preference representation variable and its structural equation\\
        $Y$, $f_Y$ & Recommendation result variable and its structural equation\\
         $Z $ & Exogenous variables in $G$\\
        $\mathbf{Z}_{u}$, $\mathbf{Z}_{v}$ & Latent vectors of exogenous variables for a user variable $u$ and an item variable $v$\\
       $\mathbf{A}_{u}$, $\mathbf{A}_{v}$ & Causal adjacency vector for a user variable $u$ and an item variable $v$\\
          $\mathbf{d}_{u}$, $\mathbf{d}_{v}$ & Feature vectors for a user variable $u$ and an item variable $v$\\
        $\mathbf{u}$, $\mathbf{v}$ & Latent factors for a user $u$ and an item $v$ \\
        $\mathbf{e}$, $\mathbf{y}$ & A user preference vector and a user interaction vector \\
        $\mathbf{h}_u$, $\mathbf{h}_v$ & Hidden factors for a user $u$ and an item $v$ from the semi-implicit generative model\\
        $\mathbf{m}_{uv}$ & Neighbor message from a node $v$ for a user $u$\\
        $\theta_1$, $\theta_2$, $\theta_3$ & Network parameters for the user encoder, the item encoder and the collaborative filtering decoder\\
        $\phi_1$, $\phi_2$ &  Network parameters for the aggregation operators\\
        $\varphi_1$, $\varphi_2$ & Network parameters for the causality-aware message passing operators\\
        $l$ & A graph learning layer \\
        $do(i=x)$ & The do-operator that forces a variable $i$ to take the value $x$\\
        \hline
    \end{tabular}
    \label{tab:notion}
\end{table*}

\noindent \textbf{Notations.}
We use uppercase letters such as $U$ to denote a set of variables. In particular, we use $U, V, E, Y$ to represent user, item,  preference representation and recommendation variables. We use lowercase letters such as $u$ to represent a random variable. In particular, we use $u, v, e, y$ to represent a specific user, item, preference and recommendation variable. Moreover, we use bold font lowercase to represent the latent vector embeddings, such as $\mathbf{u}, \mathbf{v}, \mathbf{e}, \mathbf{y} \in \mathbb{R}^d$, where $d$ is the dimension of the embedding vectors.
The weight matrix and bias vector are denoted as $\mathbf{W}$ and $\mathbf{b}$, respectively.
Primary notations are also complemented in Table~\ref{tab:notion}.

\subsection{Motivation}
\begin{definition}[Causal Graph]
\label{def:cg}
A \emph{Causal Graph}~\cite{bareinboim2022pearl} is a directed acyclic graph (DAG) $G=(\{\mathcal{V}, Z\}, \mathcal{E})$ represents causal relations among endogenous and exogenous variables.
$\mathcal{V}$ is a set of endogenous variables of interest, e.g., user and item nodes in the graph learning.
$Z$ is a set of exogenous variables outside the model, e.g., item exposure.
$\mathcal{E}$ is the edge set denoting causal relations among $G$.
\end{definition}

\begin{figure}[h]
    \centering
    \includegraphics[width=0.45\textwidth]{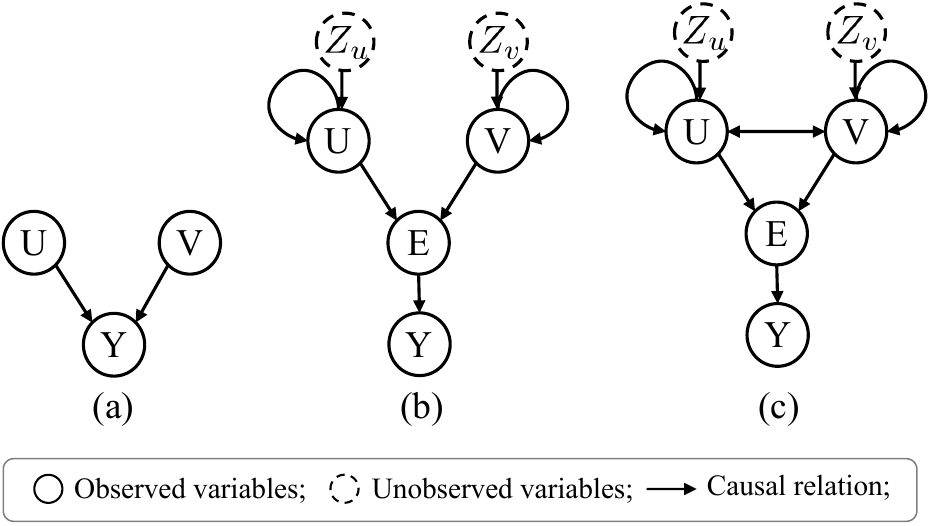}
    \caption{
    Paradigms of user preference modeling in a class of CFs:
    (a) Early CF, (b) GCF, and (c) Our causality-aware GCF.
    $Z_u$ represents hidden exogenous variables for users, e.g., user conformity; $Z_v$ are hidden exogenous variables for items, e.g., item exposure. $U$ and $V$ denote user and item, respectively. $E$ denotes preference representations from graph representation learning. $Y$ represents users' predicted recommendations.
    }
    \label{fig:scm}
\end{figure}

Following Definition~\ref{def:cg}, we start by providing the causal graphs of a class of CF methods, including early CF methods in Figure~\ref{fig:scm} (a) and existing GCF methods in Figure~\ref{fig:scm} (b).
Specifically, we aim to show the fundamental drawback shared by these two types of methods: they are fragile in capturing complex user-item relations by assuming the co-occurrence of users and items is independent.
We put forward our defined causal graph in Figure~\ref{fig:scm} (c), which considers user-item dependencies for better user preference modeling.

Early CFs largely resort to user-item associative matching~\cite{he2017neural} and follow the causal graph shown in Figure~\ref{fig:scm} (a), where user node $U$ and item node $V$ constitute a collider to affect the recommendation result $Y$.
For example, matrix factorization~\cite{mnih2007probabilistic} typically assumes $P(Y=1 \mid u, v) \propto \mathbf{u}^{\top} \mathbf{v}$, where $\mathbf{u}$ and $\mathbf{v}$ are user and item IDs and the probability of recommendations $Y$ is estimated from matching the inner product between $\mathbf{u}$ and $\mathbf{v}$. 
Latent factor-based methods~\cite{agarwal2009regression} assume $P(Y=1 \mid u, v) \propto \operatorname{LFM}(u)^{\top} \operatorname{LFM}(v)$, where $\operatorname{LFM}$ is a latent factor model that learns the user and item latent vectors, and a simple inner product is used for similarity matching to determine recommendations.

As shown in Figure~\ref{fig:scm} (b), GCF works on graph-structure data to consider auxiliary information, e.g., user/item attributes, which potentially captures exogenous variables 
$Z_u$ and $Z_v$, e.g., user conformity, item exposure.
Besides, as user-user and item-item relations are propagated through multi-hop neighbors within the graph, GCF can capture the inner connections of users and items to model more complex user behavior patterns, e.g., user collaborative behavior~\cite{wang2019neural}.
However, existing GCF methods still assume the independence between users and items. 
This is because user and item embeddings are learned separately from the graph representation learning and then subsequently applied to a CF model for user-item associative matching.
For example, NGCF~\cite{wang2019neural} assumes $P(Y=1 \mid u, v) \propto E = \operatorname{CF}\left(\operatorname{agg}\left(u, z_u, \operatorname{msg}\left(N_u\right)\right), \operatorname{agg}\left(v, z_v, \operatorname{msg}\left(N_v\right)\right)\right)$, where $\operatorname{CF}$ is a CF model for user-item associative matching. 
$N_u$ and $N_v$ are neighbor sets for users and items; $\operatorname{agg}$ and $\operatorname{msg}$ are the aggregation and message passing operations, respectively. 

In summary, both Figure~\ref{fig:scm} (a) and (b) assume the co-occurrence of users and items is independent in the observational data, i.e., there is no edge $U \to V$ or $V \to U$.
However, this assumption is unrealistic in the real world because user behaviors are influenced by the recommended items for various reasons.
For instance, users may be more likely to click the items if they are recommended~\cite{wang2018deconfounded}, which is also known as the item exposure bias~\cite{bhadani2021biases} problem.
Besides, the exposure of items is determined by user preferences estimated from the recommendation model~\cite{wang2020disentangled}, which is the essence of the personalized recommendation.
Therefore, we conceptualize the causal relations under GCN-based GCF as the Causal Graph in Figure~\ref{fig:scm} (c).
Our Causal Graph includes the modeling of $U \longleftrightarrow V$, such that user-item relations can be captured for better user preference modeling.
By given the Causal Graph in Figure~\ref{fig:scm} (c), the directed edge $\left(u\rightarrow v\right) \in \mathcal{E}$ captures the causal relation from a user $u$ to an item $v$, where $u \in U$ and $v \in V$ and $u$ is a parent node of $v$, i.e., $u \in pa\left(v\right)$.
$G$ induces a set of causal adjacency vectors $\mathbf{A}_{u}$ and $\mathbf{A}_{v}$, which specify the neighbors of a user node $u$ and an item node $v$, respectively. 
Each element $\mathbf{A}_{u}^{v} =1$ if $v \in pa(u)$, otherwise, $\mathbf{A}_{u}^{v}=0$.
Similarly, $\mathbf{A}_{v}^{u}=1$ if $u \in pa(v)$.

\subsection{Formulation}
The key innovation of this work is to integrate causal modeling into the learning process of a GCN-based GCF model. 
The problem can be formulated as,
\begin{definition}[Problem Formulation]
Establish the connection between the GCN-based GCF model and the Causal Graph depicted in Figure~\ref{fig:scm} (c).
Motivated by Neural-Causal Connection~\cite{xia2021causal}, the goal is to approximate a Neural Causal Model (NCM) based on the provided Causal Graph.
\end{definition}
To achieve this goal, we first convert the Causal Graph into a Structural Causal Model (SCM) (Section~\ref{sec:SCM}). 
Subsequently, the NCM is defined based on the SCM, with each structural equation in the SCM corresponding to a neural network in the NCM (Section~\ref{sec:NCM}).
To approximate the trainable neural networks in the NCM, we employ a unified learning framework described in Section~\ref{sec:VAE}. 
This framework enables causal modeling, making the learned graph embeddings as expressive as the causal effects modeled by the SCM.
Overall, through the integration with causal modeling, our approach offers a novel perspective on graph representation learning, leveraging the expressive power of the causality-aware graph embeddings to capture complex causal relations in the recommendation.

\section{Neural Causal Model}
This section evokes the concept of the Structural Causal Model (SCM) and the Neural Causal Model (NCM).
The SCM converts causal relations among the Causal Graph in Figure~\ref{fig:scm} (c) as structural equations;
The NCM defines each of the structural equations as a parameterized neural network.

\subsection{Structural Causal Model}\label{sec:SCM}

The Causal Graph in Figure~\ref{fig:scm} (c) has four variables of interest (i.e., endogenous variables): $U$ (user),
$V$ (item), $E$ (preference representation) and $Y$ (recommendation).
Besides, two exogenous variables $Z_u$ and $Z_v$ are manifest, representing hidden impacts such as user conformity~\cite{zheng2021disentangling} and item exposure~\cite{li2022causal}.
The causal mechanism of modeling the four endogenous variables $\{U, V, E, Y\}$ is done by a SCM~\cite{pearl2000models}.

\begin{definition}[Structural Causal Model]
\label{def:SCM}
A \emph{Structural Causal Model (SCM)}~\cite{pearl2000models} $\mathcal{M} = \langle\mathcal{V}, Z, \mathcal{F}, P(Z)\rangle$ is the mathematical form of the Causal Graph $G$ that includes a collection of structural equations $\mathcal{F}$ on endogenous variables $\mathcal{V}$ and a distribution $P(Z)$ over exogenous variables $Z$.
A structural equation $f_{U} \in \mathcal{F}$ for a variable $u \in U \subseteq \mathcal{V}$ is a mapping from $u$'s parents and exogenous variables of $u$:
\begin{equation}
    u \leftarrow f_{U}\left(pa(u), Z_u\right), Z_u \sim P(Z)
\end{equation}
where $pa(u) \subseteq \mathcal{V} \backslash u$ is $u$'s parents from the Causal Graph $G$.
$Z_u \in Z$ is a set of exogenous variables connected with $u$.
\end{definition}
Following Definition~\ref{def:SCM} and the causal relations in Figure~\ref{fig:scm} (c), endogenous variables $\{U, V, E, Y\}= \mathcal{V}$ are modeled by structural equations $\{f_U, f_V, f_E, f_Y\} = \mathcal{F}$. Formally,
\begin{equation}
\label{eq:structual}
\mathcal{F}(\mathcal{V}, Z):= \left\{\begin{array}{l}
U \leftarrow f_U\left(U, V, Z_u\right) \\
V \leftarrow f_V\left(U,V, Z_v\right) \\
E \leftarrow f_E(U, V) \\
Y \leftarrow f_Y(E)
\end{array}\right.
\end{equation}
These structural equations model the direct causal relation from a set of causes (e.g., $pa(u)$) to a variable (e.g., $u \in U$) accounting for the impact of exogenous variables (e.g., $Z_u$).

\subsection{Neural Network for Causal Model}\label{sec:NCM}

We now formally introduce Neural-Causal Connection~\cite{xia2021causal}, i.e., the connection between deep neural networks (e.g., GCNs) and causal models is done by establishing an NCM.
\begin{definition}[Neural-Causal Connection]
\label{def:Neural-Causal}
A \emph{Neural Causal Model} (NCM)~\cite{xia2021causal} is defined as $\mathcal{M}(\theta)$ and is parameterized for the SCM $\mathcal{M}$ in Definition~\ref{def:SCM}. 
Each structural equation in $\mathcal{M}$ is defined as a feedforward neural network in $\mathcal{M}(\theta)$, e.g., Multi-layer perceptron (MLP).
Exogenous variables $Z$ are mapped into hidden vectors ${\mathbf{Z}}$ that follow the Gaussian distribution $\mathcal{N}\left(0, \mathbf{I}_K\right)$.
\end{definition}

The NCM $\mathcal{M}(\theta)$ is expressive~\cite{xia2021causal}, such that it generates distributions associated with the Pearl Causal Hierarchy (PCH)~\cite{pearl2018book}, i.e., modeling ``observational'' (layer 1), ``interventional'' (layer 2) and ``counterfactual'' (layer 3) distributions.

In accordance with Definition~\ref{def:Neural-Causal}, we aim to build an NCM $\mathcal{M}(\theta)$ that models structural equations defined in Eq.~\eqref{eq:structual} as parameterized feedforward neural networks. 
Formally, 
\begin{equation}
\label{eq:GNCM}
 \mathcal{M}(\theta) \triangleq  \left\{\begin{array}{l}
\mathbf{Z}_u \sim \mathcal{N}\left(0, \mathbf{I}_K\right), \mathbf{Z}_v \sim \mathcal{N}\left(0, \mathbf{I}_K\right), \\ 
\mathbf{u} \propto f_U = q_{\theta_1}(f_{\phi_1}\left(\mathbf{Z}_u, f_{\varphi_1}(U \mid U, V)\right)), \\ 
\mathbf{v} \propto f_V = q_{\theta_2}(f_{\phi_2}\left(\mathbf{Z}_v, f_{\varphi_2}(V \mid U, V)\right)),\\
\mathbf{e} \propto f_E =p_{\theta_3}\left(\mathbf{u}, \mathbf{v}\right), \\ 
\mathbf{y}\sim f_Y = \operatorname{Multinomial}\left(N, \mathbf{e}\right)   \\
\end{array}\right.
\end{equation}
\begin{itemize}
    \item $Z_u$, $Z_v$ are mapped into low-dimensional hidden vectors $\mathbf{Z}_u$ and $\mathbf{Z}_v$ using Gaussian distribution $\mathcal{N}\left(0, \mathbf{I}_K\right)$.
    \item $\mathbf{u} \propto f_U$: user representation $\mathbf{u}$ is calculated by a user encoder $q_{\theta_1}$.
    The user encoder takes as input the aggregated (i.e., $f_{\phi_1}$) information of user exogenous variables $\mathbf{Z}_u$ and user's causality-aware neighbor messages $f_{\varphi_1}$.
    \item $\mathbf{v} \propto f_V$: item representation $\mathbf{v}$ is given by an item encoder $q_{\theta_2}$.
    The item encoder uses aggregated (i.e., $f_{\phi_2}$) information of item exogenous variables $\mathbf{Z}_v$ and item's causality-aware neighbor messages $f_{\varphi_2}$.
    \item $\mathbf{e} \propto f_E$: 
    user preference probability $\mathbf{e}$ is produced by a collaborative filtering decoder $p_{\theta_3}$ by using latent representations $\mathbf{u}$ and $\mathbf{v}$.
    \item $\mathbf{y}\sim f_Y$: 
    user interaction $\mathbf{y}$ is sampled from a multinomial distribution with the probability $\mathbf{e}$.
    $N$ is the user's total interaction number.    
\end{itemize}

\section{Variational Inference for NCGCF}
\label{sec:VAE}

\begin{figure*}
    \centering
    \includegraphics[width=0.95\textwidth]{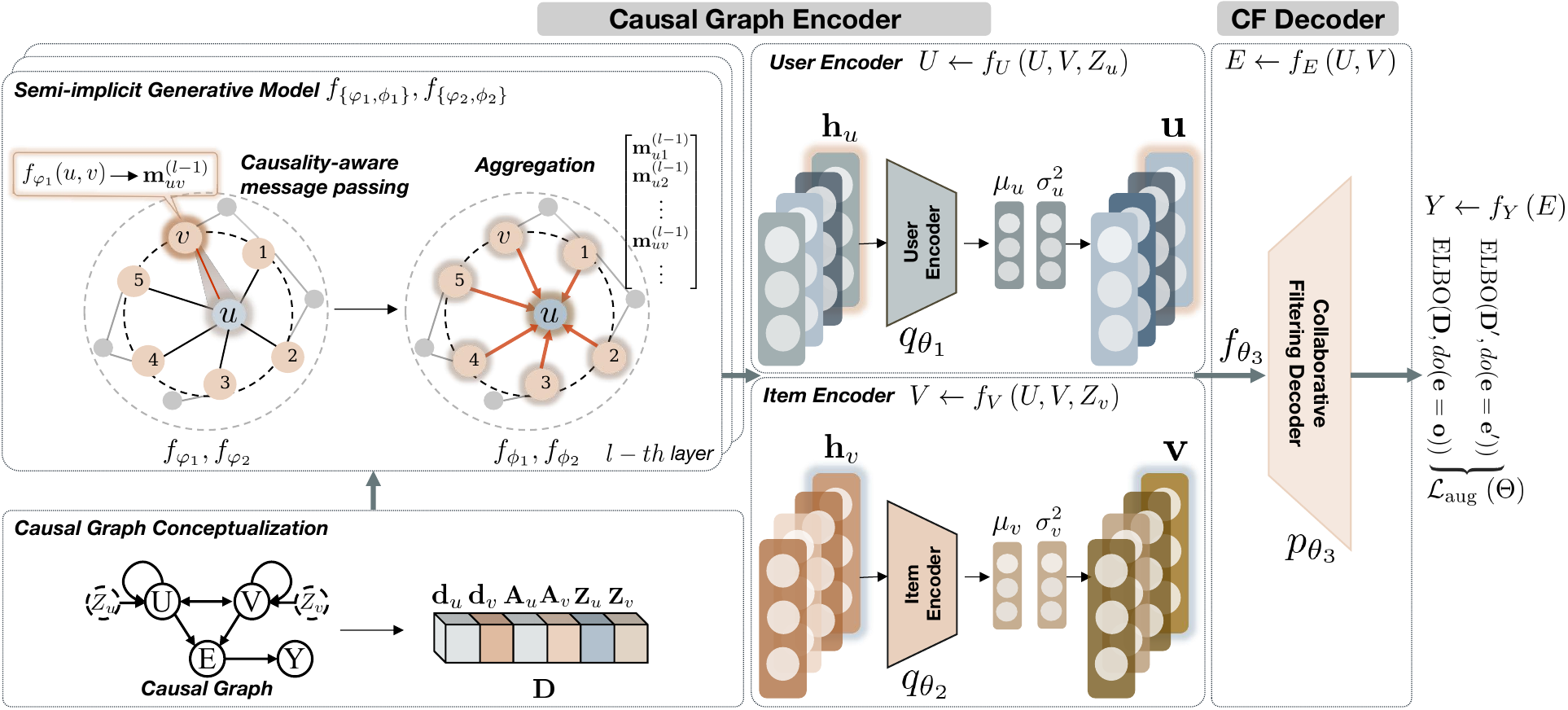}
    \caption{NCGCF framework: causal graph conceptualization prepossess a user-item interaction graph by using the causal relations under our defined Causal Graph; causal graph encoder models the causal relations under the graph-structured data using a semi-implicit generative model, and outputs user and item representations with a user encoder and an item encoder;  collaborative filtering (CF) decoder uses CF to construct preference vectors based on user and item representations. Finally, NCGCF is optimized through a counterfactual instance-aware ELBO to capture user preference shifts.}
    \label{fig:framework}
    \vspace{-0.25in}
\end{figure*}

We now introduce our framework, namely, \emph{Neural Causal Graph Collaborative Filtering (NCGCF)}.
We show the NCGCF framework in Figure~\ref{fig:framework}, which includes three major components based on the variational autoencoder structure:
\begin{itemize}
    \item Causal Graph Encoder: approximates $f_U$ and $f_V$.
The causal graph encoder includes a user encoder, an item encoder and a semi-implicit generative model.
The semi-implicit generative model calculates causal relations between nodes as causality-aware messages.
The user encoder and item encoder then use these causality-aware messages to output user representation and item representation, respectively.
    \item Collaborative Filtering (CF) Decoder: 
approximates $f_E$ using a CF method to estimate user preference.
    \item Counterfactual Instances-based Optimization: optimizes model parameters by implementing $f_Y$ with counterfactual instances to capture user preference shifts.
\end{itemize}

\subsection{\textbf{Causal Graph Encoder}}
The causal graph encoder aims to model $f_U$ and $f_V$ in Eq.~\eqref{eq:GNCM}.
However, this is not a trivial task as the true posteriors of $f_U$ and $f_V$ do not follow standard Gaussian distributions due to the existence of causal relations between node pairs.
Besides, these causal relations should be modeled into causality-aware messages using neural networks. 
Thus, traditional variational inference~\cite{kipf2016variational} that approximates posteriors to simple, tractable Gaussian vectors is not applicable.
Semi-implicit variational inference (SIVI)~\cite{yin2018semi} that models complex distributions through implicit posteriors proves to be an effective alternative~\cite{hasanzadeh2019semi,8588399}.
Inspired by SIVI, we devise a semi-implicit generative model on top of the user and item encoder to model implicit posteriors.
In particular, the semi-implicit generative model calculates causal relations between nodes as causality-aware messages.
Those causality-aware messages are encoded into user and item hidden factors $\mathbf{h}_u$ and $\mathbf{h}_v$.
Then, the user encoder takes $\mathbf{h}_u$ as the input to output the user representation $\mathbf{u}$.
Analogously, the item encoder uses $\mathbf{h}_v$ to calculate item representation.

\subsubsection{\textbf{Semi-implicit Generative Model}}
contains two operators, namely, \emph{causality-aware message passing} and \emph{aggregation}.
The \emph{causality-aware message passing} uses learnable neural networks to model each of the dependency terms for a node and its neighbors within a structural equation.
For example, $f_{\varphi_1}(u,v)$ models the dependency between a user node $u$ and his/her neighbor item node $v$, such that the learned message becomes a descriptor of the causal relation $\left(u \rightarrow v\right)$. 
The \emph{aggregation} uses weighted-sum aggregators to aggregate user/item exogenous variables and the calculated causality-aware neighbor messages. 
Finally, user and item hidden factors $\mathbf{h}_u$ and $\mathbf{h}_v$ are output for latter user and item encoder learning.
\begin{itemize}
\item \emph{Causality-aware message passing}: 
For the user encoder, given user $u$'s features $\mathbf{d}_u$ and its causal adjacency vector $\mathbf{A}_u$, the messages from $u$'s neighbor $v$ is given by:
\begin{equation}
\label{eq:message}
\begin{gathered}
    \mathbf{m}_{uv}^{(l-1)} = f_{\varphi_1}(u,v)= \operatorname{MLP}^{(l)}\left( \mathbf{h}_u^{(l-1)} \| \mathbf{h}_v^{(l-1)}\right)
\\    
=\operatorname{ReLU}\left(\mathbf{W}_{\varphi_1}^{(l)}\left( \mathbf{h}_u^{(l-1)}\| \mathbf{h}_v^{(l-1)}\right)\right), \text{ for } l \in \{1, \cdots, L\}
\end{gathered}
\end{equation}
where $\mathbf{m}_{uv}^{(l-1)}$ is the neighbor message at the $l-1$-th graph learning layer~\footnote{The neighbor message at the $0$-th layer, i.e., $\mathbf{m}_{uv}^{(0)}$, is initialized from a normal distribution.}.
$v$ is a neighbor for $u$ and $v \in N_u \propto \mathbf{A}_u$.
$\mathbf{h}_v^{(l-1)}$ and $\mathbf{h}_u^{(l-1)}$ are hidden factors for the neighbor $v$ and the user $u$ at the $l-1$-th layer~\footnote{$\mathbf{h}_v^{(0)}$ and $\mathbf{h}_u^{(0)}$ are initialized as node features $\mathbf{d}_v$ and $\mathbf{d}_u$.}.
$\mathbf{W}_{\varphi_1}^{(l)}$ is the weight matrix for $f_{\varphi_1}$ at the $l$-th layer and $\|$ denotes column-wise concatenation. 
Analogously, for the item encoder, we can calculate the neighbor message $\mathbf{m}_{vu}^{(l-1)}$ for an item $v$ by replacing $f_{\varphi_1}$ with $f_{\varphi_2}$ in Eq.~\eqref{eq:message}.
\end{itemize}

\begin{itemize}
\item \emph{Aggregation}:
For the user encoder, at each graph learning layer $l$, we perform aggregation operation on the messages $\mathbf{m}_{uv}^{(l-1)}$ from $u$'s neighbors and the user exogenous variables $\mathbf{Z}_u$ to obtain the hidden factor $\mathbf{h}_u^{(l)}$:
\begin{equation}
\label{eq:aggreation}
\mathbf{h}_u^{(l)}=\left(\mathbf{h}_u^{(l-1)} \| f_{\phi_1}\left(\left\{\mathbf{W}_{\phi_1}^{(l)} \mathbf{m}_{uv}^{(l-1)}: v \in N_u\right\}\right)  \| \mathbf{Z}_u \right)
\end{equation}
where $\mathbf{h}_u^{(l)}$ is the learned hidden factor for $u$ at the $l$-th graph learning layer.
$f_{\phi_1}$ is the aggregation operator chosen as weighted-sum, following~\cite{altaf2019dataset}. $\mathbf{W}_{\phi_1}^{(l)}$ is the weight for $f_{\phi_1}$ that specifies the different contributions of neighbor messages to the target node at the $l$-th layer. $\|$ is the column-wise concatenation. 
$\mathbf{Z}_u$ is low-dimensional latent factors for user exogenous variables given by Gaussian distribution $\mathcal{N}\left(0, \mathbf{I}_K\right)$.
Similarly, for the item encoder, we calculate item $v$'s hidden factors $\mathbf{h}_v^{(l)}$ by using $f_{\phi_2}$ with $\mathbf{W}_{\phi_2}$ in Eq.~\eqref{eq:aggreation}.
\end{itemize}

Having obtained the hidden factors $\mathbf{h}_u^{(l)}$ for user $u$ and $\mathbf{h}_v^{(l)}$ for item $v$ at each graph learning layer $l \in \{1,\cdots, L\}$, we adopt layer-aggregation~\cite{xu2018representation} to concatenate vectors at all layers into a single vector:
\begin{equation}
\label{eq:hidden semi}
\mathbf{h}_u=\mathbf{h}_u^{(1)} + \cdots + \mathbf{h}_u^{(L)}, \quad
\mathbf{h}_v=\mathbf{h}_v^{(1)} + \cdots + \mathbf{h}_v^{(L)}
\end{equation}
By performing layer aggregation, we capture higher-order connectivities of node pairs across different graph learning layers.
Finally, our semi-implicit generative model outputs $\mathbf{h}_u$ and $\mathbf{h}_v$ as hidden factors of users and items.

\subsubsection{\textbf{User and Item Encoder}}
Given hidden factors $\mathbf{h}_u$ for a user $u$, the user encoder outputs mean and variance in $\mathcal{N}\left(\mu_u, \operatorname{diag}\left(\sigma_u^2\right)\right)$, from which user embedding $\mathbf{u}$ is sampled:
\begin{equation}
\label{eq:user encoder}
    q_{\theta_1}\left(\mathbf{u} \mid \mathbf{h}_u\right) =\mathcal{N}\left(\mathbf{u} \mid \mu_u, \operatorname{diag}\left(\sigma_u^2\right)\right)  
\end{equation}
where $\mu_u$ and $\operatorname{diag}\left(\sigma_u^2\right)$ are the mean and variance for user $u$, which are obtained by sending $u$'s hidden factors $\mathbf{h}_u$ to a one-layer neural network with the activation function $\operatorname{ReLU}(x)=\max (0, x)$:
\begin{equation}
\mu_u=\operatorname{ReLU}\left(\mathbf{W}^{\mu_u}_{\theta_1} \mathbf{h}_u+\mathbf{b}\right),  \sigma_u^2=\exp \left(\operatorname{ReLU}\left(\mathbf{W}^{\sigma_u}_{\theta_1} \mathbf{h}_u+\mathbf{b}\right)\right)
\end{equation}
where $\mathbf{W}_{\theta_1} = \{\mathbf{W}^{\mu_u}_{\theta_1}, \mathbf{W}^{\sigma_u}_{\theta_1}  \}$ is a hidden-to-output weight matrix for the user encoder $q_{\theta_1}$; $\mathbf{b}$ is the bias vector.
Analogously, the item encoder follows the same paradigm as the user encoder to generate the mean and variance for item $v$ based on $v$'s hidden factors $\mathbf{h}_v$:
\begin{equation}
\label{eq:item encoder}
\begin{gathered}
        q_{\theta_2}\left(\mathbf{v} \mid \mathbf{h}_v\right) =\mathcal{N}\left(\mathbf{v} \mid \mu_v, \operatorname{diag}\left(\sigma_v^2\right)\right),  \\\mu_v=\operatorname{ReLU}\left(\mathbf{W}^{\mu_v}_{\theta_2} \mathbf{h}_v+\mathbf{b}\right), \sigma_v^2=\exp \left(\operatorname{ReLU}\left(\mathbf{W}^{\mu_v}_{\theta_2} \mathbf{h}_v+\mathbf{b}\right)\right)
\end{gathered}
\end{equation}
where $\mathbf{W}_{\theta_2} = \{\mathbf{W}^{\mu_v}_{\theta_2}, \mathbf{W}^{\sigma_v}_{\theta_2}  \}$ is the weight matrix for the item encoder $q_{\theta_2}$.

\subsection{\textbf{Collaborative Filtering Decoder}} 
Collaborative filtering is largely dominated by latent factor models, as evidenced by Koren et al.~\cite{koren2008factorization}. These models involve mapping users and items into latent factors in order to estimate the preference scores of users towards items.
We use latent factor-based collaborative filtering in our decoder for modeling the user preference $\mathbf{e}$, which is a probability vector over the entire item set for recommendations. 
The predicted user interaction vector $\mathbf{y}$ is assumed to be sampled from a multinomial distribution with probability $\mathbf{e}$.

Formally, we define a generative function $f_{\theta_3}(\mathbf{u}, \mathbf{v})$ recovering classical latent factor-based CF to approximate user preference vector $\mathbf{e}$:
\begin{equation}\label{eq:uv}
    \mathbf{e} = \operatorname{softmax}(f_{\theta_3}(\mathbf{u}, \mathbf{v})) = \operatorname{softmax}(\mathbf{u}^\top\mathbf{v})
\end{equation}
where $\mathbf{u}$ and $\mathbf{v}$ are latent factors for a user $v$ and an item $v$ drawn from Eq.~\eqref{eq:user encoder} and Eq.~\eqref{eq:item encoder}, respectively. The $\operatorname{softmax}$ function transforms the calculated preference scores to probability vector $\mathbf{e}$ over the item corpus.

Then, the decoder $p_{\theta_3}\left(\mathbf{e} \mid \mathbf{u}, \mathbf{v} \right)$ produces interaction probability $\mathbf{y}$ by approximating a logistic log-likelihood:
\begin{equation}
\log p_{\theta_3}\left(\mathbf{y} \mid \mathbf{e}\right) = 
    \sum_v y_{uv} \log \sigma\left(\mathbf{e}\right)+\left(1-y_{uv}\right) \log \left(1-\sigma\left(\mathbf{e}\right)\right)
\end{equation}
where $y_{uv}$ is the historical interaction between $u$ and $v$, e.g., click. $\sigma(\mathbf{e})=1 /(1+\exp (-\mathbf{e}))$ is the logistic function. 

\subsection{\textbf{Counterfactual Instances-based Optimization}} 

We wish our NCGCF to be robust to unseen (unknown) user preference shifts to further enhance the recommendation robustness.
Catching user preferences is at the core of any recommendation model~\cite{koren2022advances}; however, user preferences may change over time~\cite{10.1145/3485447.3512072,wang2022mgpolicy}.  
For example, a user may once love items with the brand \textit{Nike} but change his taste for liking \textit{Adidas}.
Such a user preference shift can be captured by actively manipulating user preferences, i.e., manipulating $\mathbf{e}$.

Since our NCGCF is a Neural Causal Model and is capable of generating ``interventional'' distributions (cf. Section~\ref{sec:NCM}) within the Pearl Causal Hierarchy, the manipulations can be done by performing interventions~\cite{bareinboim2022pearl} on the user preference vector $\mathbf{e}$ using a do-operator $do(\cdot)$, i.e., $do(\mathbf{e}= \mathbf{e}^{\prime})$.
The data after interventions are called \emph{counterfactual instances}~\cite{xiong2021counterfactual} that, if augmented to original training instances, increase the model robustness to unseen interventions (i.e., user preference shifts). 
Inspired by~\cite{zhang2020causal}, we optimize NCGCF by considering two data scenarios, i.e., the clean data scenario in which our NCGCF accesses the data without interventions, and the counterfactual data scenario in which the data is generated by known interventions on user preference vectors.

Formally, for the clean data scenario, assuming that NCGCF observes only clean data $\mathbf{D}$ during training. 
In this case, we retain the original value $\mathbf{o}$ of user preference $\mathbf{e}$ by using $do(\mathbf{e}=\mathbf{o})$.
Then, NCGCF is trained by maximizing the likelihood function $ \log p_{\theta_3}\left(\mathbf{y} \mid do(\mathbf{e}=\mathbf{o})\right) $.
Since this marginal distribution is intractable~\cite{kipf2016variational,liang2018variational}, we instead maximize the intervention evidence lower-bound (ELBO) with $do(\mathbf{e}=\mathbf{o})$, i.e. $\max_{\theta_1, \theta_2,\theta_3} \operatorname{ELBO}(\mathbf{D}, do(\mathbf{e}=\mathbf{o}))$.
In particular, 
\begin{equation}
\begin{aligned}
    & \operatorname{ELBO}(\mathbf{D}, do(\mathbf{e}=\mathbf{o}))
     = \\
    & \mathbb{E}_{\theta} 
    \left[\log \frac{p_{\theta_3}\left(\mathbf{y} \mid  do(\mathbf{e}=\mathbf{o} \right) ) p(\mathbf{u})p(\mathbf{v})}{q_{\theta_1}\left(\mathbf{u} \mid \Xi, do(\mathbf{e}  =\mathbf{o}) \right)q_{\theta_2}\left(\mathbf{v} \mid \Xi, do(\mathbf{e}=\mathbf{o}) \right)}\right] \\
     = &\mathbb{E}_{\theta} \left[\log p_{\theta_3}\left(\mathbf{y} \mid do(\mathbf{e}=\mathbf{o} \right) )\right] \\
   &   - \operatorname{KL}\left(q_{\theta_1}\left(\mathbf{u} \mid \Xi\right) \| p\left(\mathbf{u}\right), q_{\theta_2}\left(\mathbf{v} \mid \Xi\right) \| p\left(\mathbf{v}\right)\right)
\end{aligned}
\end{equation}
where $\Xi$ represents required parameters for the conditional probability distributions of $q_{\theta_1}$, $q_{\theta_2}$ and $p_{\theta_3}$, i.e., $\Xi =\{\mathbf{Z}_u, \mathbf{d}_u, \mathbf{A}_u\}$ for $q_{\theta_1}$, 
$\Xi =\{\mathbf{Z}_v, \mathbf{d}_v, \mathbf{A}_v\}$ for $q_{\theta_2}$ and $\Xi =\{\mathbf{u}, \mathbf{v} \}$ for $p_{\theta_3}$.
$\theta=\{\theta_1, \theta_2, \theta_3\}$ is a set of model parameters and $\operatorname{KL}(\cdot )$ is KL-divergence between two distributions.

For the counterfactual data scenario, we assume NCGCF accesses counterfactual data $\mathbf{D}^{\prime}$ generated by known interventions $do(\mathbf{e}=\mathbf{e}^{\prime})$ on user preference vectors.
The counterfactual vectors $\mathbf{e}^{\prime}$ hold the same dimension with $\mathbf{e}$ and are drawn from a random distribution. 
Then, the ELBO of NCGCF with the counterfactual data is,
\begin{equation}
\begin{aligned}
    \operatorname{ELBO} & (\mathbf{D}^{\prime}, do(\mathbf{e}=\mathbf{e}^{\prime}))
    =\mathbb{E}_{\theta} \left[\log p_{\theta_3}\left(\mathbf{y} \mid do(\mathbf{e}=\mathbf{e}^{\prime} \right) )\right] \\
    & -\operatorname{KL}\left(q_{\theta_1}\left(\mathbf{u} \mid \Xi\right) \| p\left(\mathbf{u}\right), q_{\theta_2}\left(\mathbf{v} \mid \Xi\right) \| p\left(\mathbf{v}\right)\right)
\end{aligned}
\end{equation}

Inspired by data augmentation and adversarial training~\cite{van2001art}, we augment the clean data with counterfactual instances to enhance the robustness of our NCGCF meanwhile capturing user preference shifts. 
In particular, the total loss function after augmentation is as below,
\begin{equation}
\begin{aligned}
    \mathcal{L}_{\text {aug }}(\Theta)&=\lambda(\operatorname{ELBO}(\mathbf{D}, do(\mathbf{e}=\mathbf{o})) \\
    &+(1-\lambda) (\operatorname{ELBO}(\mathbf{D}^{\prime}, do(\mathbf{e}=\mathbf{e}^{\prime}))
 \end{aligned}   
\end{equation}
where $\mathcal{L}_{\text {aug }}(\Theta)$ is the loss function for training our NCGCF and $\Theta$ are model parameters. $\lambda$ is the trade-off parameter between the clean and the counterfactual data scenario.
During the training stage, the loss function is calculated by averaging the ELBO over all users. 

\section{Experiments}
We thoroughly evaluate the proposed NCGCF for the recommendation task to answer the following research questions:
\begin{itemize}
    \item \textbf{RQ1:} 
    How does NCGCF perform as compared with state-of-the-art recommendation methods? 
    \item \textbf{RQ2:} How do different components impact NCGCF's performance? 
    \item \textbf{RQ3:} How do parameters in the causal graph encoder affect NCGCF?
    
\end{itemize}

\subsection{Experimental Settings}
\subsubsection{Datasets}
We conduct experiments on one synthetic dataset and three real-world datasets to evaluate the effectiveness of NCGCF.
The synthetic dataset is constructed in accordance with the Causal Graph depicted in Figure~\ref{fig:scm}(c).
The construction process follows a series of assumptions that reflect causal relations between users and items. 
For instance, we assume the causal relation from user features to user preferences based on prior knowledge, such as a positive effect of high income on the preference over high price.
Similar assumptions also apply to item features to user preferences, e.g., the positive effect of the brand ``Apple'' on the preference for high-priced items.
In particular, the \textbf{Synthetic} dataset construction is under the following four steps:
\begin{enumerate}
    \item Feature generation:
    We simulate $|U|=1,000$ users and $|I|=1,000$ items, where each user has one discrete feature (\textit{gender}) and one continuous feature (\textit{income}), while each item has three discrete features, i.e., \textit{type}, \textit{brand} and \textit{price}. 
    For discrete features, their values in $\{0,1\}$ are sampled from Bernoulli distributions.
    We sample continuous features from random sampling, in which random feature values are chosen from the minimum (i.e., $0$) and the maximum (i.e., $1000$) feature values. 
    For both users and items, we assume two exogenous variables (i.e., $Z_u$ and $Z_v$) drawn from the Gaussian distribution.
    \item 
    Causal neighbor sampling: 
    We synthesize the causal relations $U \to U$ and $V \to V$ by creating user/item causal neighbors.
    In particular, we set the causal neighbor number $N_c=10$.
    We assume a user $u$'s causal neighbors ($U \to U$) are those who have interacted with the same item with the user $u$. 
    In other words, users who have shown interest in similar items are considered causal neighbors for each other.
    For item causal neighbor sampling ($V \to V$), we first convert items with their features into dense vectors through item2vec~\cite{barkan2016item2vec}, then calculate the Euclidean distances between two items. 
    We assume those items that have the $N_c$ smallest distances from the target item are causal neighbors for the target item. 
    \item 
    User preference estimation:
    For each user $u$ and item $v$, the user preference $\mathbf{u} \in \mathbb{R}^d$ towards item property $\mathbf{v} \in \mathbb{R}^d$ is generated from a multi-variable Gaussian distribution $\mathcal{N}(0, \mathbf{I})$.
    Then, the preference score $y_{uv}$ between user $u$ and item $v$ is calculated by the inner product of $\mathbf{u}$ and $\mathbf{v}$.
    Besides, we assume the fine-grained causal relations from user/item features to the preference score based on prior knowledge. 
    For example, we assume a positive effect of the ``high'' \textit{income} on the preference over ``high'' \textit{price}, thus tuning the preference score to prefer items with high prices. 
    Besides, a user should have similar preference scores toward an item and the item's causal neighbors.
    \item User interaction sampling:
    Once we obtain a user $u$'s preference scores for all items (i.e., $I$), we normalize preference scores by $\frac{\exp \left(r_{i}\right)}{\sum_{i^\prime \in I} \exp \left(r_{i^{\prime}}\right)}$.
    We select items with $k$-top scores as the interactions for the user $u \in U$, where $k$ is a constant chosen randomly from range $[20, 100]$.
\end{enumerate}

Apart from the synthetic dataset, we also use three benchmark datasets to test our performance in real-world scenarios.
We also assume fine-grained causal relations in these real-world datasets to ensure users interact with items causally.
\begin{itemize}
    \item 
    \textbf{Amazon-Beauty} and \textbf{Amazon-Appliances}: two sub-datasets from Amazon Product Reviews~\footnote{https://nijianmo.github.io/amazon/index.html}~\cite{he2016ups}, which record large crawls of user reviews and product metadata (e.g., \textit{brand}). 
    Following~\cite{haque2018sentiment}, we use \textit{brand} and \textit{price} to build item features since other features (e.g., \textit{category}) are too sparse and contain noisy information. 
    We use co-purchased information from the product metadata to build item-item causal relations, i.e., $V \to V$.
    The co-purchased information records item-to-item relationships, i.e., a user who bought item $v$ also bought item $i$.
    We assume an item's causal neighbors are those items that are co-purchased together.
    For user-user causal relation (i.e., $U \to U$), we assume a user's causal neighbors are those who have similar interactions, i.e., users who reviewed the same item are neighbors for each other. 
    \item \textbf{Epinions}~\footnote{http://www.cse.msu.edu/~tangjili/trust.html}~\cite{tang-etal12a}: 
    a social dataset recording social relations between users.
    We convert user/item features from the dataset into one-hot embeddings.
    We use social relations to build user causal neighbors, i.e., a user's social friends are the neighbors of the user. 
    Besides, items bought by the same user are causal neighbors to each other. 
\end{itemize}

For the three real-world datasets, we regard user interactions with overall ratings above $3.0$ as positive interactions.
For the synthetic dataset, we regard all user-item interactions as positive as they are top items selected based on users' preferences.
We adopt a $10$-core setting, i.e., retaining users and items with at least ten interactions.
The statistics of the four datasets are shown in Table~\ref{tb:dataset}.
For model training, we randomly split samples in both datasets into training, validation, and test sets by the ratio of 70\%, 10\%, and 20\%.

\begin{table}[h]
\centering
\caption{Statistics of the datasets. 
}\label{tb:dataset}
\resizebox{0.5\textwidth}{!}{
\begin{tabular}{l|c|c|c|c}
\hline
\multicolumn{1}{c|}{Dataset}& {Synthetic} & {Amazon-Beauty} & {Amazon-Appliances} & {Epinions}   \\
\hline
\# Users & 1,000  & 271,036 & 446,774  & 116,260
\\ \# Items & 1,000  & 29,735   & 27,888  & 41,269
\\ \# Interactions & 12,813  & 311,791   & 522,416  & 181,394
\\ \# Density & 0.0128  & 0.0039   & 0.0041  & 0.0038
\\
\hline
\end{tabular}
}
\end{table}

\subsubsection{Baselines}
 We compare NCGCF with eight competitive recommendation methods.
\begin{itemize}
    \item \textbf{BPR}~\cite{rendle2012bpr}: a well-known matrix factorization-based model with a pairwise ranking loss to enable recommendation learning from implicit feedback.
    \item \textbf{NCF}~\cite{he2017neural}: extends CF to neural network architecture. It maps users and items into dense vectors and feeds user and item vectors into an MLP to predict user preferences.
    \item \textbf{MultiVAE}~\cite{liang2018variational}: extends CF to variational autoencoder (VAE) structure for implicit feedback modeling. 
    It formulates CF learning as a generative model and uses variational inference to model the posterior distributions.
    \item \textbf{NGCF}~\cite{wang2019neural}: a GCF that incorporates two GCNs to learn user and item embeddings. The learned embeddings are passed to a matrix factorization to capture the collaborative signal for recommendations.
    \item \textbf{VGAE}~\cite{kipf2016variational}: a graph learning method that extends VAE to handle graph-structured data. We use VGAE to obtain user and item embeddings and inner product those embeddings to predict user preference scores.
    \item \textbf{GC-MC}~\cite{berg2017graph}: a graph-based auto-encoder framework for matrix completion. The encoder is a GCN that produces user and item embeddings. The learned embeddings reconstruct the rating links through a bilinear decoder.
    \item \textbf{LightGCN}~\cite{he2020lightgcn}: a SOTA graph-based recommendation model that simplifies the GCN component.
    It includes the essential part in GCNs, i.e., neighbor aggregation, to learn user and item embeddings for collaborative filtering.
    \item \textbf{CACF}~\cite{zhang2021causally}: a method that learns attention scores from individual treatment effect estimation.
    The attention scores are used as user and item weights to enhance the CF. 
\end{itemize}

\subsubsection{Evaluation Metrics}
We use three Top-$K$ recommendation evaluation metrics, i.e., Precision@$K$, Recall@$K$ and Normalized Discounted Cumulative Gain(NDCG)@$K$. 
The three evaluation metrics measure whether the recommended Top-$K$ items are consistent with users' preferences in their historical interactions. 
We report the average results with respect to the metrics over all users.
The Wilcoxon signed-rank test~\cite{doi:https://doi.org/10.1002/9780471462422.eoct979} is used to evaluate whether the improvements against baselines are significant.

\subsubsection{Parameter Settings}
We implement our NCGCF~\footnote{\url{https://github.com/Chrystalii/CNGCF}} using Pytorch.
The latent embedding sizes of neural networks for all neural-based methods are fixed as $d=64$.
The in-dimension and out-dimension of the graph convolutional layer in NCGCF, NGCF, VGAE, GC-MC and LightGCN is set as $32$ and $64$, respectively for graph learning. 
We apply a dropout layer on top of the graph convolutional layer to prevent model overfitting for all GCN-based methods.  
The Adam optimizer is applied to all methods for model optimization, where the batch size is fixed as 1024.
The hyper-parameters of all methods are chosen by the grid search, including the learning rate $l_r$ in $\{0.0001,0.0005,0.001,0.005\}$, $L_2$ norm regularization in $\left\{10^{-5}, 10^{-4}, \cdots, 10^1, 10^2\right\}$, and the dropout ratio $p$ in $\{0.0,0.1, \cdots, 0.8\}$.
We set the maximum epoch for all methods as $400$ and use the early stopping strategy, i.e., terminate model training when the validation Precision@10 value does not increase for 20 epochs.
To ensure a fair comparison, all baseline methods are trained using the same data used in our NCGCF. 
This includes using causality-enhanced node features and causal relations, such as item-item and user-user relationships, in the training process for all models.

\subsection{Recommendation Performance (RQ1)}

\begin{table*}[htbp]
\caption{Recommendation performance comparison: The best results are highlighted in bold while the second-best ones are underlined.
All improvements against the second-best results are significant at $p < 0.01$.
}\label{tab:overall}
\centering
\setlength{\tabcolsep}{5pt}
\resizebox{\textwidth}{!}{
\begin{tabular}
    {c ccc ccc ccc ccc} \toprule
    Dataset
    & \multicolumn{3}{c}{{Synthetic}} 
    & \multicolumn{3}{c}{{Amazon-Beauty}} 
    & \multicolumn{3}{c}{ {Amazon-Appliances}}
    & \multicolumn{3}{c}{{Epinions}}\\\cmidrule(lr){2-4} \cmidrule(lr){5-7} \cmidrule(lr){8-10} \cmidrule(lr){11-13}
Method & Precision@10 & Recall@10 & NDCG@10  & Precision@10 & Recall@10 & NDCG@10  & Precision@10 & Recall@10 & NDCG@10  & Precision@10 & Recall@10 & NDCG@10 \\\midrule

\midrule

BPR &0.5214&0.4913&0.6446  &0.3555 &0.3319 &0.4111  & 0.3720  & 0.3574  & 0.4356   &0.3022&0.2895	&0.4889 \\
NCF &0.6120 &0.6293&0.7124  &0.3618 &0.3659 &0.4459  & 0.3871  & 0.3789  & 0.4771  &0.3551&0.3364 &0.5432\\
MultiVAE &0.6248&0.5999&0.8101  &0.4418 &0.4112 &0.4616 &0.4544  & 0.4428  & 0.5998 &0.4229&0.3888 &0.5331\\
NGCF  &0.5990 &0.5681&0.7477  &0.4512 &0.4003 &0.5188  &0.4271  & 0.3778  & 0.5555  &0.4018&0.3912&0.5012\\
VGAE  &0.5446&0.5572&0.7778  &0.3499&0.3812&0.4466  &0.3681  & 0.4014  & 0.5019  &0.3590&0.3460 & 0.4913\\
GC-MC &0.6115&0.6226&0.8116  &0.4666 &0.4615 &\underline{0.5612}  &0.4718  & 0.4518  & 0.5677  &0.4666&0.4218 &0.5112\\
LightGCN &\underline{0.6439}&\underline{0.6719} &\underline{0.8223}  &\underline{0.4810} &\underline{0.4778}&0.5501 &\underline{0.4844} & \underline{0.4652}  & \underline{0.6028}   &\underline{0.4717}&\underline{0.4544}& \underline{0.5436}\\
CACF &0.4482&0.4158&0.5555  &0.3101&0.3005&0.3888 &0.3222&0.3188 &0.4215 &0.2899&0.2765&0.3445\\
\textbf{NCGCF}  &\textbf{0.7952}  &\textbf{0.6889}  &\textbf{0.8538}  & \textbf{0.5148}  & \textbf{0.5183}  & \textbf{0.6855}  & \textbf{0.6510}  & \textbf{0.5271}  &\textbf{0.8193}  &\textbf{0.4990 } &\textbf{0.5030}  &\textbf{0.5589} \\
\textbf{Improv.\%} &+23.4\%  &+2.5\%  &+3.8\%  &+7.0\%  &+8.4\%  & +22.1\% &+34.3\%  &+13.3\%  &+35.9\%  & +5.7\% &+10.6\%  &+2.8\% \\

\midrule
& Precision@20 & Recall@20 & NDCG@20  & Precision@20 & Recall@20 & NDCG@20  & Precision@20 & Recall@20 & NDCG@20  & Precision@20 & Recall@20 & NDCG@20 \\\midrule

\midrule
BPR &0.6111&0.5536&0.6338  &0.3561&0.3420&0.4062
& 0.3941 & 0.3599 & 0.4322
&0.3332&0.3232&0.4689 \\
NCF &0.6678&0.6446&0.7003 &0.3699&0.3691&0.4330 & 0.3999  & 0.4033 & 0.4519
&0.3719&0.3614&\underline{0.5255}\\
MultiVAE &0.6779&0.6136&0.8006  &0.4496	&0.4200	&0.4555 & 0.4819 & 0.4716  & \underline{0.5911} &0.4465&0.4055 &0.5133\\
NGCF &0.6233&0.5999&0.7312 &0.4612&0.4112&0.5081  & 0.4666  & 0.4258  & 0.5499&0.4223&0.4210&0.4811\\
VGAE  &0.5847&0.5687&0.7613  &0.3551&0.3999&0.4410
& 0.3771 & 0.4228 & 0.4761&0.3667&0.3598&0.4781 \\
GC-MC  &0.6665&0.6317&0.8091 &0.4781&0.4771&\underline{0.5582}
& 0.4892 & \underline{0.4881}  & 0.5514 &0.4815&0.4451&0.4999 \\
LightGCN  &\underline{0.6904} &\underline{0.6819}&\underline{0.8108} &\underline{0.5023}&\underline{0.4869}&0.5306 & \underline{0.4919} & 0.4781 & 0.5613 &\underline{0.4915}&\underline{0.4718}	&0.5221 \\
CACF &0.4567&0.4266&0.5348  &0.3186&0.3211&0.3678 &0.3418&0.3271	&0.4103 &0.2747&0.2910 &0.3368\\
\textbf{NCGCF}  &\textbf{0.8081}  &\textbf{0.6844 } &\textbf{0.8603}  & \textbf{0.5153}  &\textbf{0.5106}  &\textbf{0.7123}  &\textbf{0.6367}  &\textbf{0.5055 } &\textbf{0.8501}  &\textbf{0.5002}  &\textbf{0.5034}  &\textbf{0.5667}\\
\textbf{Improv.\%} &+17.0\%  &+0.3\%  &+6.1\%  & +2.5\% &+4.8\%  &+27.6\%  &+29.4\%  &+3.5\%  &+43.8\%  &+1.7\%  &+6.6\%  &+7.8\%\\

\bottomrule
\end{tabular}
}
\vspace{-0.25in}
\end{table*}

We show the recommendation performance of our NCGCF and all baselines on the four datasets in Table~\ref{tab:overall}. 
By analyzing Table~\ref{tab:overall}, we have the following findings.
\begin{itemize}
\item NCGCF consistently outperforms the strongest baselines on both synthetic and real-world datasets, achieving the best recommendation performance across all three evaluation metrics.
In particular, NCGCF outperforms the strongest baselines by 23.4\%, 7.0\%, 34.3\% and 5.7\% in terms of Precision@10 on Synthetic, Amazon-Beauty, Amazon-Appliances and Epinions, respectively.
Additionally, NCGCF improves Recall@10/NDCG@10 by 2.5\%/3.8\%, 8.4\%/22.1\%, 13.3\%/35.9\% and 10.6\%/2.8\% on the four datasets, respectively.
The superiority of NCGCF can be attributed to two factors: the power of neural graph learning and the modeling of causality. 
Firstly, graph learning explicitly models the interactions between users and items as a graph, and uses graph convolutional networks to capture the non-linear relations from neighboring nodes.
This allows graph learning to capture more complex user behavior patterns.
Secondly, modeling causal relations allows us to identify the causal effects of different items on users, thus capturing true user preferences on items.
By injecting causal modeling into graph representation learning, our NCGCF captures more precise user preferences to produce robust recommendations against baselines. 
    
\item 
NCGCF achieves the most notable improvements (e.g., 35.9\% for NDCG@10 and 43.8\% for NDCG@20) on the Amazon-Appliances dataset, which is a large-scale dataset with a considerable amount of user behavior data that may be noisy and challenging to model.
NCGCF's ability to inject causality into graph learning enables the model to surpass merely capturing spurious correlations among noisy data, leading to more accurate and reliable modeling of true user preferences.

\item NGCF that uses graph representation learning outperforms NCF without graph learning.
This is because NGCF models user-item interactions as a graph, and uses graph convolutional networks to capture more complex user-user collaborative behavior to enhance recommendations.
In contrast, NCF uses a multi-layer perception to learn user and item similarities, which captures only linear user-item correlations from the interaction matrix.
Moreover, GC-MC and LightGCN outperform other graph learning-based baselines (i.e., NGCF, VGAE) in most cases.
This is because GC-MC and LightGCN aggregate multiple embedding propagation layers to capture higher-order connectivity within the interaction graph.
Similarly, our NCGCF incorporates layer aggregation within our causal graph encoder, enabling us to capture higher-order connectivity and produce better graph representations for improved recommendation performance.

\item NCGCF outperforms all graph learning-based baselines, including NGCF, VGAE, GC-MC and LightGCN.
This is because NCGCF models causal relations within the graph learning process.
Guided by the causal recommendation generation process, NCGCF is able to inject causal relations under the Structural Causal Model into the learning process of the graph convolutional network. 
This allows NCGCF to uncover the causal effect of items on users and capture user behavior patterns more accurately.

\end{itemize}
    
\subsection{Study of NCGCF (RQ2)}

\begin{table}[htbp]
\centering
\caption{Recommendation performance after replacing the causal graph encoder with different graph representation learning methods. 
The value after $\pm$ indicates the increase or decrease of the variant's performance compared with NCGCF.
}\label{tb:ablation_gcns}
 \resizebox{0.5\textwidth}{!}{
\begin{tabular}{ c  c c c }
\hline
Variants & Precision@10 & Recall@10 & NDCG@10
\\ \hline
& \multicolumn{3}{c}{{Synthetic}} \\ \hline
NCGCF & 0.7952  & 0.6889  & 0.8538 \\  
NCGCF-{GCN} &0.5358(-32.7\%) &0.5182(-24.7\%)  &0.7025( -17.7\%)  \\
NCGCF-{Graphsage} & 0.5038(-36.8\%) &0.5005(-27.4\%)  &0.7022(17.8\%) \\
NCGCF-{Pinsage} &0.5819(-26.8\%) & 0.5498(-20.2\%) &0.7446( -12.8\%)  \\

\hline
& \multicolumn{3}{c}{{Amazon-Beauty}} \\\hline
NCGCF & 0.5148  & 0.5183  & 0.6855\\  
NCGCF-{GCN} & 0.4991(-3.04\%) & 0.5029(-2.97\%)  & 0.4886(-28.68\%)  \\
NCGCF-{Graphsage} &0.5011(-2.67\%) & 0.5039(-2.78\%)  & 0.5243(-23.55\%) \\
NCGCF-{Pinsage} &0.5008(-2.72\%) &0.5043(-2.70\%)  &0.5143(-25.01\%)   \\

\hline
& \multicolumn{3}{c}{{Amazon-Appliances}} \\\hline
NCGCF & 0.6510  & 0.5271  &0.8193 \\  
NCGCF-{GCN} &0.5067(-3.04\%) &0.5167(-2.97\%) &0.6614(-28.68\%)  \\
NCGCF-{Graphsage} &0.5085(-2.67\%) &0.5184(2.78\%)  &0.6670(- 23.55\%)\\
NCGCF-{Pinsage} &0.5083(-2.72\%) & 0.5178(-2.70\%)  &0.6631(-25.01\%)   \\

\hline
& \multicolumn{3}{c}{{Epinions}} \\\hline
NCGCF &0.4990  &0.5030  &0.5589 \\  
NCGCF-{GCN} & 0.4812(-3.55\%) &0.4990(-0.79\%) &0.5013(-10.28\%) \\
NCGCF-{Graphsage} &0.4809(-3.62\%) &0.4989(-0.81\%)  & 0.4999(-10.52\%)  \\
NCGCF-{Pinsage} &0.4871(-2.38\%) &0.4994(-0.71\%)  &0.4930(-11.74\%)  \\

\hline
\end{tabular}}
\end{table}

We start by exploring how replacing our causal graph encoder with other graph representation learning methods, i.e., naive GCN~\cite{kipf2016semi}, Graphsage~\cite{hamilton2017inductive} and Pinsage~\cite{ying2018graph}, impact NCGCF's performance. 
We then analyze the influences of core components, including causality-aware message passing and counterfactual instance-aware ELBO.

\subsubsection{\textbf{Effect of Causal Graph Encoder}}
The causal graph encoder plays a pivotal role in NCGCF to model the causal relations of nodes.
To investigate its effectiveness, we replace our causal graph encoder with different encoders built by other graph learning methods.
In particular, we use GCN~\cite{kipf2016semi}, Graphsage~\cite{hamilton2017inductive} and Pinsage~\cite{ying2018graph} to produce user and item embedding vectors for the decoder learning phase, and compare the performance of NCGCF before and after the replacements.
We present the experimental results in Table~\ref{tb:ablation_gcns}.
We find that both GCN~\cite{kipf2016semi}, Graphsage~\cite{hamilton2017inductive} and Pinsage~\cite{ying2018graph}-based encoders downgrade the performance of NCGCF compared with NCGCF equipped with our proposed causal graph encoder. 
For instance, NCGCF with a GCN-based encoder downgrades the NDCG@10 by 28.68\% on the Amazon-Beauty. 
This is because GCN, Graphsage and Pinsage cannot capture the causal relations of nodes in the interaction graph, leading to insufficient representations of users and items.
On the contrary, our causal graph encoder captures the intrinsic causal relations between nodes using the causality-aware message passing; thus, it learns causality-aware user and item representations to better serve the later decoder learning. 
Moreover, the GCN-based encoder downgrades the NCGCF performance most severely compared with GraphSage and Pinsage-based encoders.
This is because naive GCN performs transductive learning requiring full graph Laplacian, whereas GraphSage and Pinsage perform inductive learning without requiring full graph Laplacian to handle large-scale graph data well. 
We thus conclude that an inductive learning setting is more desired for our NCGCF, especially when facing large-scale graph data.

\begin{table}[htbp]
\centering
\caption{Ablation Study on NCGCF. 
$\neg$ CM represents causality-aware message passing is removed.
$\neg$ CI represents counterfactual instance-aware ELBO is removed.
}\label{tb:ablation}
 \resizebox{0.45\textwidth}{!}{
\begin{tabular}{ c c c c }
\hline
Variants & Precision@10 & Recall@10 & NDCG@10
\\ \hline
& \multicolumn{3}{c}{{Synthetic}} \\\hline
NCGCF  &0.7952  &0.6889  &0.8538\\  
$\neg$ CM  & 0.5806($-31.9\%$) & 0.5491($-20.3\%$)  & 0.7179($-16.0\%$)   \\  
$\neg$ CI &0.7781($-2.1\%$) & 0.6654($-3.4\%$)  & 0.7573($-11.2\%$)  \\ 

\hline
& \multicolumn{3}{c}{{Amazon-Beauty}} \\\hline
NCGCF & 0.5148  & 0.5183  & 0.6855 \\  
$\neg$ CM &0.5007($-2.7\%$) &0.5060($-2.3\%$)  &0.5383($-20.7\%$)   \\  
$\neg$ CI &0.5101($-0.9\%$) &0.5081($-2.0\%$)  & 0.5738($-15.9\%$)  \\ 

\hline
& \multicolumn{3}{c}{{Amazon-Appliances}} \\\hline
NCGCF & 0.6510  & 0.5271  &0.8193 \\  
$\neg$ CM &0.6357($-2.4\%$) &0.5050($-4.2\%$)  &0.6864($-16.2\%$) \\  
$\neg$ CI &0.6445($-1.0\%$) & 0.5143($-2.4\%$) &0.7956($-2.9\%$)   \\ 

\hline
& \multicolumn{3}{c}{{Epinions}} \\\hline
NCGCF &0.4990  &0.5030  &0.5589 \\  
$\neg$ CM &0.4695($-6.0\%$) & 0.4936($-1.9\%$) &0.4647($-16.9\%$)  \\  
$\neg$ CI &0.4794($-3.9\%$) &0.5018($-0.2\%$) &0.5139($-8.1\%$) \\ 
\hline
\end{tabular}}
    \vspace{-0.25in}
\end{table}

\subsubsection{\textbf{Effect of Causality-aware Message Passing}}
The causality-aware message passing models the dependency terms between each of the structural equations as the causal relations between nodes.
We present NCGCF's performance after removing the causality-aware message passing in Table~\ref{tb:ablation}.
We observe that removing the component downgrades NCGCF's performance, indicating the importance of causality-aware message passing in helping NCGCF to achieve favorable recommendation performance. 
We thus conclude that modeling the causal relations between nodes within the graph-structured data is essential for graph learning-based models to uncover true user preferences for improved recommendations.

\subsubsection{\textbf{Effect of Counterfactual Instance-aware ELBO}}  
The counterfactual instance-aware ELBO augments counterfactual instances for NCGCF optimization.
We present NCGCF's performance after removing the counterfactual instance-aware ELBO in Table~\ref{tb:ablation}.
Apparently, removing the counterfactual instance-aware ELBO leads to the downgraded performance of NCGCF on both datasets.
This is because our counterfactual instance-aware ELBO augments counterfactual instances, i.e., the intervened data on user preference vectors, thus facilitating better model optimization to capture user preference shifts.

\subsection{Parameter Analysis of Causal Graph Encoder (RQ3)}

\begin{figure}[!h]
\centering
\begin{minipage}[t]{0.24\textwidth}
\centering
\includegraphics[width=\textwidth]{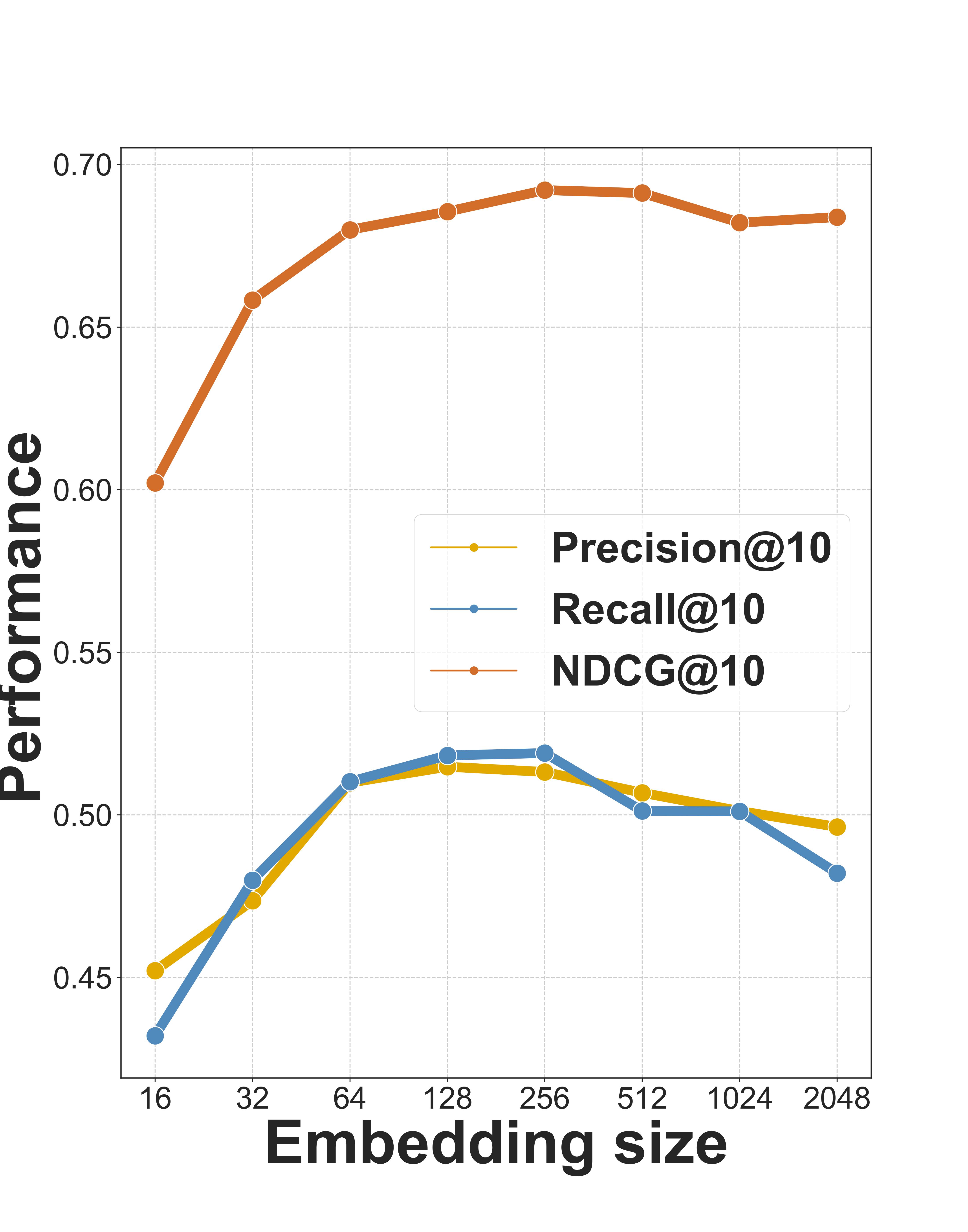}
\subcaption{Impact of embedding size on Amazon-Beauty.}
\end{minipage}
\begin{minipage}[t]{0.24\textwidth}
\centering
\includegraphics[width=\textwidth]{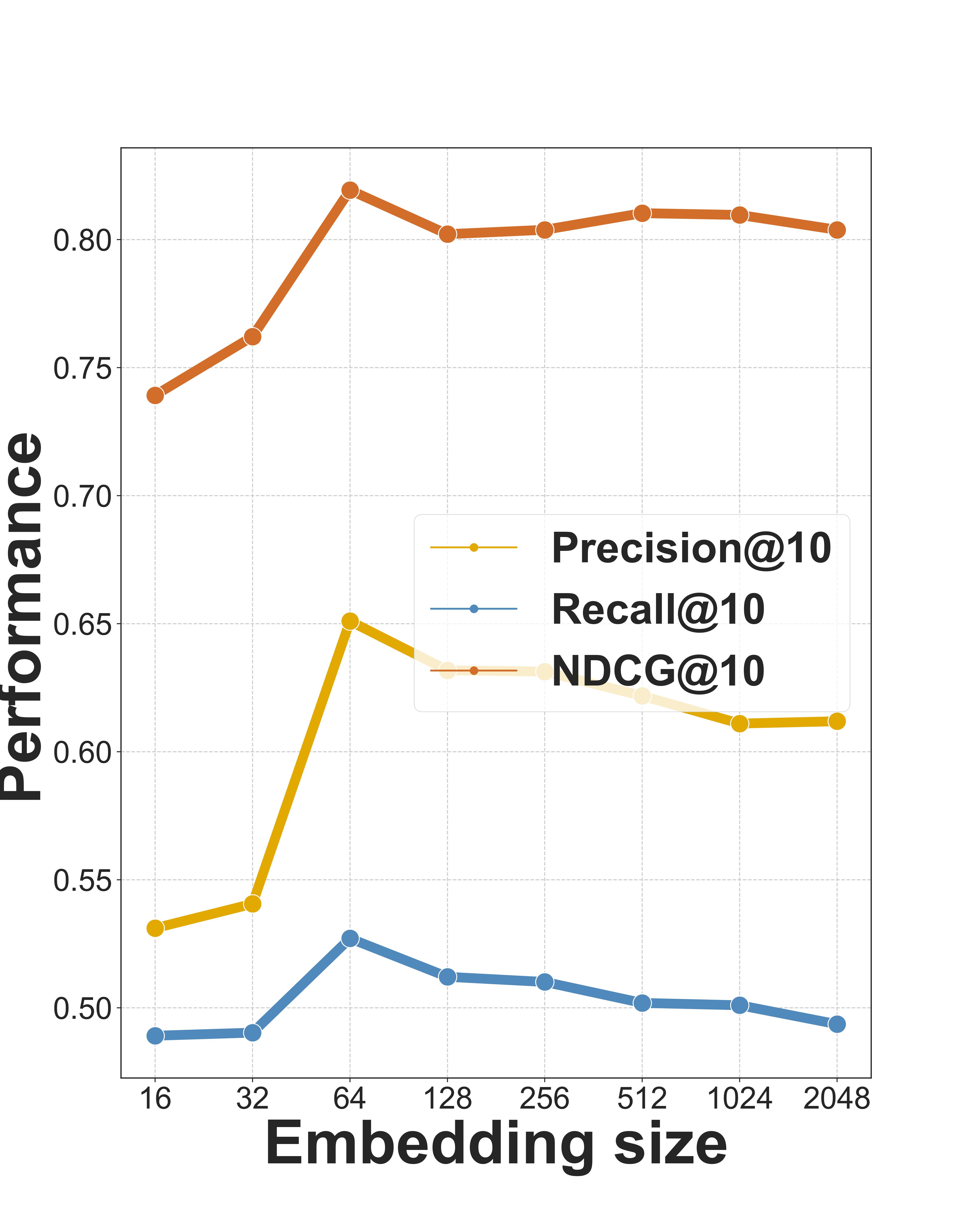}
\subcaption{Impact of embedding size on Amazon-Appliances.}
\end{minipage}
\begin{minipage}[t]{0.24\textwidth}
\centering
\includegraphics[width=\textwidth]{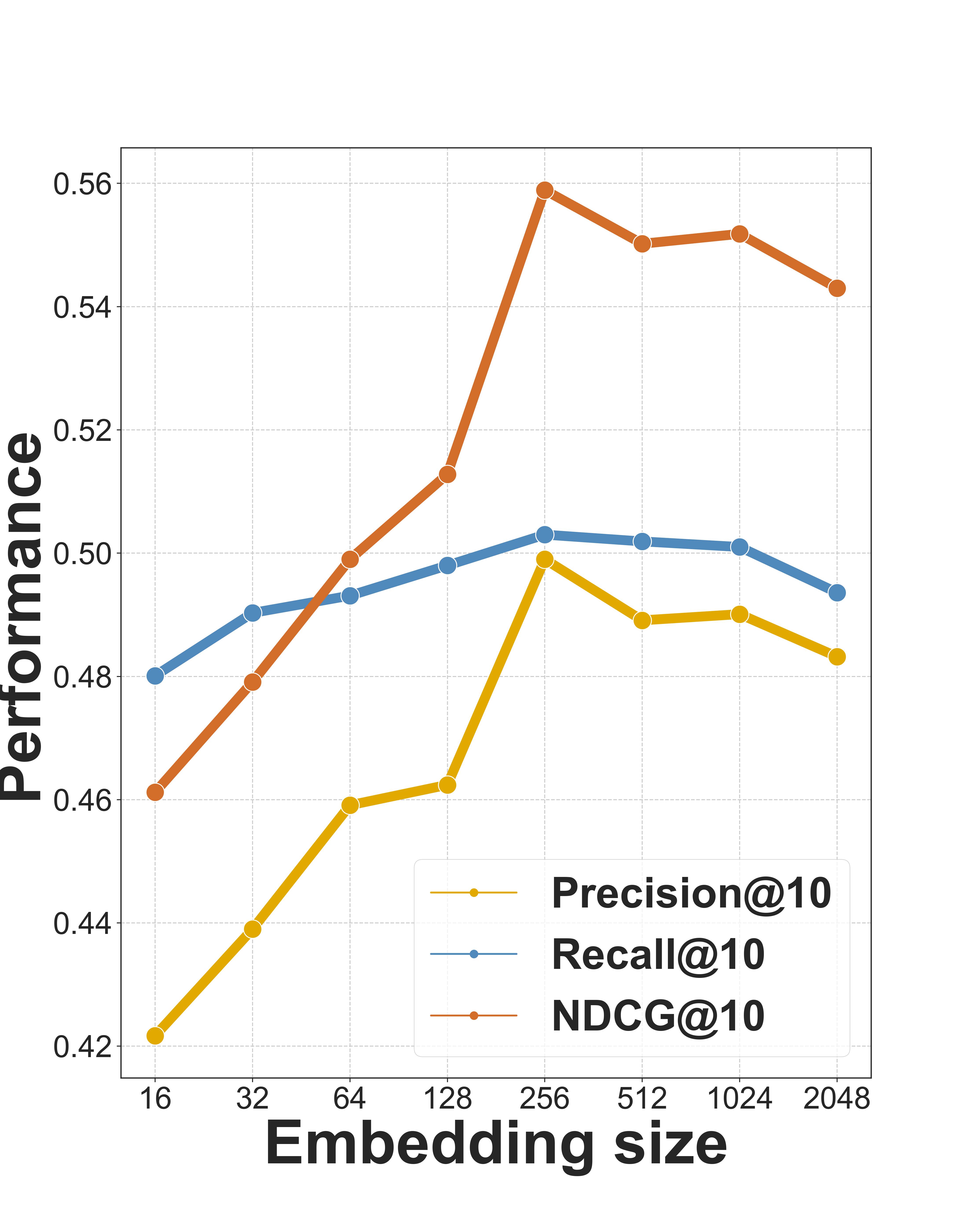}
\subcaption{Impact of embedding size on Epinions.}
\end{minipage}
\begin{minipage}[t]{0.24\textwidth}
\centering
\includegraphics[width=\textwidth]{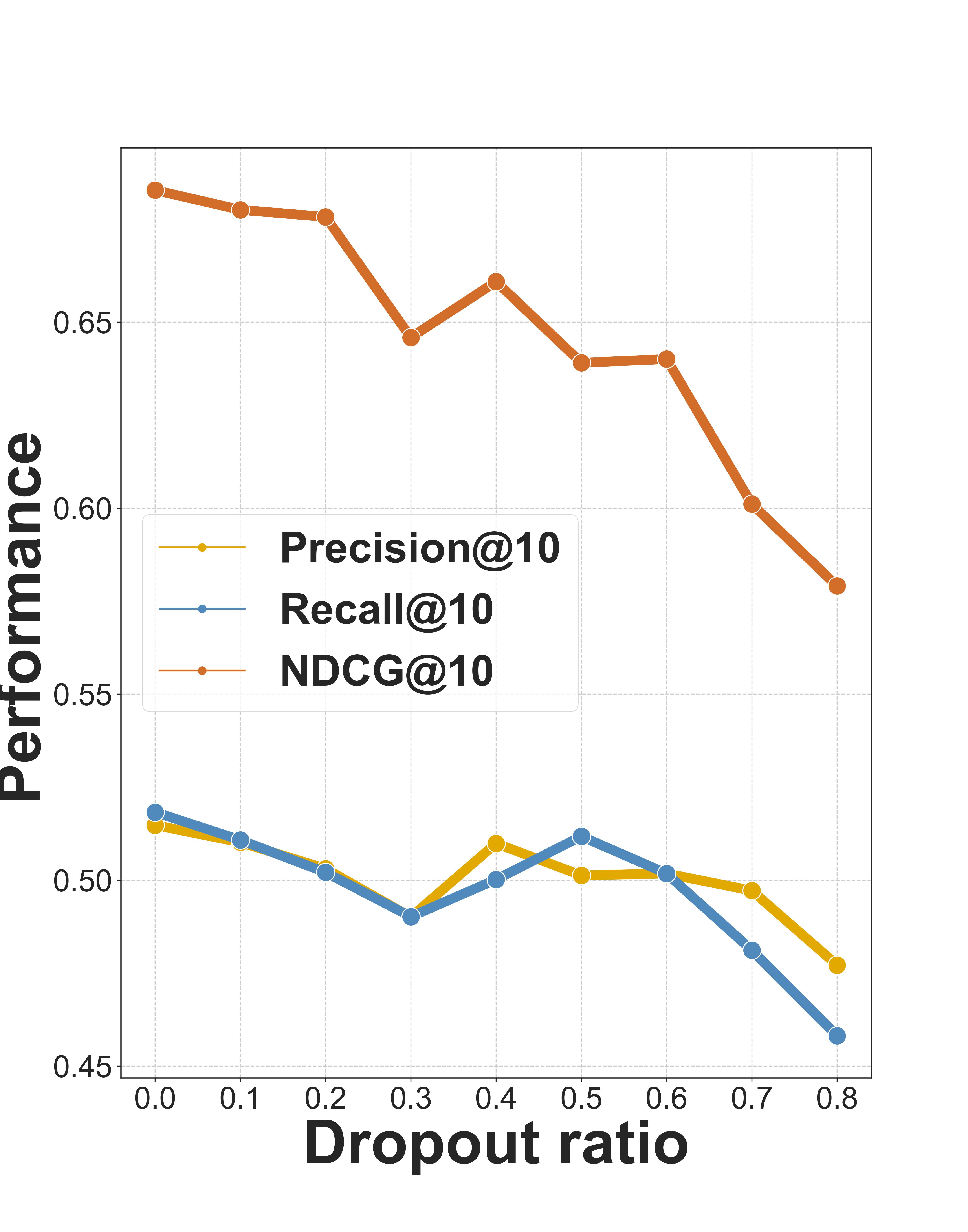}
\subcaption{Impact of dropout ratio on Amazon-Beauty.}
\end{minipage}
\begin{minipage}[t]{0.24\textwidth}
\centering
\includegraphics[width=\textwidth]{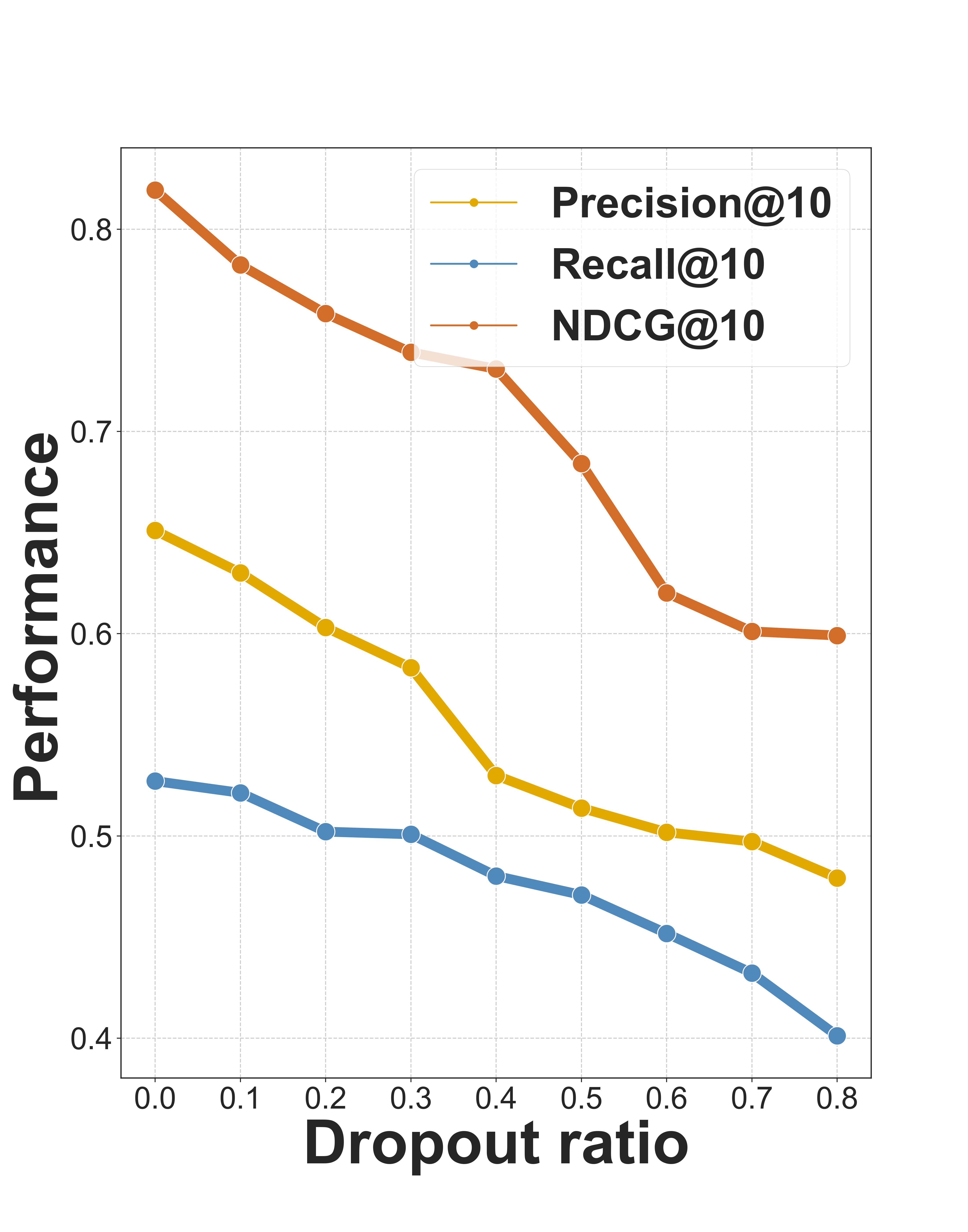}
\subcaption{Impact of dropout ratio on Amazon-Appliances.}
\end{minipage}
\begin{minipage}[t]{0.24\textwidth}
\centering
\includegraphics[width=\textwidth]{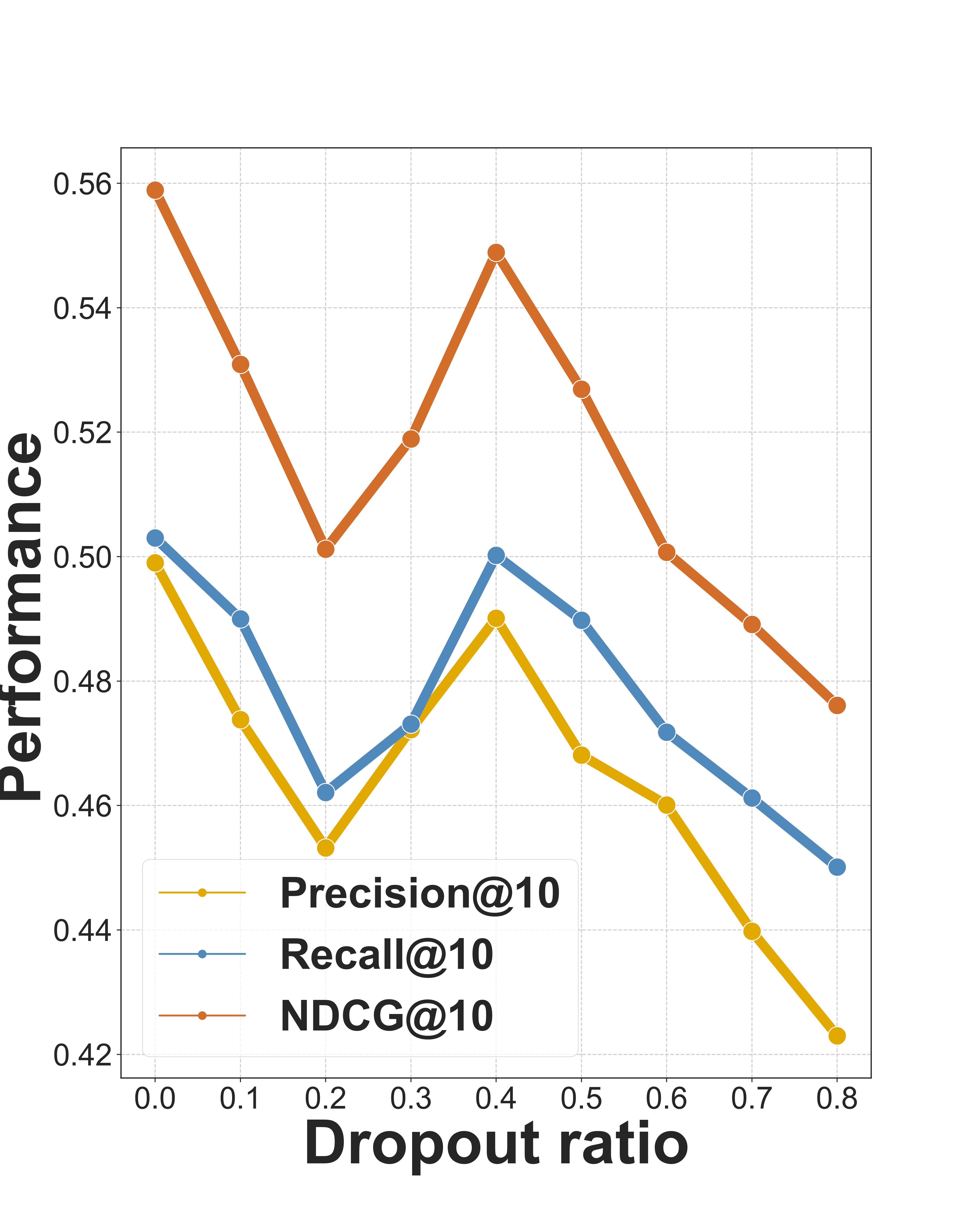}
\subcaption{Impact of dropout ratio on Epinions.}
\end{minipage}
\caption{Parameter analysis on causal graph encoder.}
\label{fig:parameter}
\vspace{-0.3in}
\end{figure}







We analyze NCGCF's performance under different embedding sizes $n$ of the semi-implicit generative model in the causal graph encoder.
We also investigate the node dropout ratios $p$ of the dropout layer applied in the causal graph encoder. 

\subsubsection{\textbf{Effect of Embedding Size}}
Figure~\ref{fig:parameter} (a) (b) (c) report the parameter sensitivity of our NCGCF w.r.t. embedding size $n$ with $n = \{16, 32, 64, 128, 256, 512, 1024, 2048\}$.
Apparently, the performance of NCGCF on Amazon-Beauty, Amazon-Appliances and Epinions demonstrates increasing trends from $n=16$, then reaches the peak when $n = 512$, $n = 64$ and $n=256$, respectively. 
This is reasonable since $n$ controls the number of latent vectors of users and items from the semi-implicit generative model, and low-dimensional latent vectors cannot retain enough information for the encoder learning phrase. 
After reaching the peaks, the performance of NCGCF degrades slightly and then becomes stable.
The decrease in performance is due to the introduction of redundant information as the embedding size becomes too large, which can affect the model. 
Additionally, we observe the largest Amazon-Appliances dataset requires the smallest embedding size of $n = 64$ to reach its peak performance compared to the other two datasets.
This is because a larger embedding size brings large-scale datasets a higher computational burden, thus limiting the model's performance.

\subsubsection{\textbf{Effect of Dropout Ratio}}
We employ a node dropout layer in the causal graph encoder to prevent model overfitting.
We show the influence of node dropout ratio $p$ on the three datasets in Figure~\ref{fig:parameter} (d) (e) (f).
We observe that the performance of NCGCF on both Amazon-Beauty, Amazon-Appliances and Epinions exhibits a decreasing trend as we increase the node dropout ratio $p$ from $0.0$ to $0.3$, but recovers at $p=0.4$. 
After $p=0.4$, the performance of NCGCF decreases as the dropout ratio increases.
We believe that the reduced performance could be attributed to the removal of crucial information that the model needs to learn from the data, thus impairing the NCGCF's performance. 
Nevertheless, the recovered performance at $p=0.4$ indicates that NCGCF is robust to balance the loss of information and overfitting.



\section{Related Work}

\subsection{Graph Collaborative Filtering}

Collaborative filtering (CF)~\cite{schafer2007collaborative} dominates recommendation research due to its simplicity and effectiveness. 
Early CF models are largely latent factor models~\cite{agarwal2009regression}.
They use descriptive features (e.g., IDs) to calculate user similarities, assuming that users with similar historical behaviors have similar future preferences.
For example, Bayesian personalized ranking (BPR)~\cite{rendle2012bpr} learns
user and item latent vectors from the interaction matrix built by implicit user feedback, e.g., clicks.
The inner products between latent vectors are used as user-item similarities to predict user preference scores.

With the burgeoning of neural models, various neural networks are used for better user preference modeling.
Neural collaborative filtering (NCF)~\cite{he2017neural} uses a Multi-layer perceptron (MLP) to learn a user behavior similarity function based on simple user/item one-hot encodings.
Recently, benefiting from the capability to learn from relational graphs, graph CF (GCF) leverages advances in graph learning~\cite{xia2021graph} to model user-item interaction graphs as well as rich auxiliary data (e.g., text, image), thus boosting the recommendation by augmenting complex semantics under user-item interactions.
Early GCF relies on random walk models to calculate similarities among users and items from the given graph.
With the rise of graph neural networks, recent GCF methods have shifted towards graph representation learning. 
Graph Convolutional Network (GCN) is one of the most wildly adopted graph neural networks for scrutinizing complex graph relations as user and item embeddings.
Neural graph collaborative filtering (NGCF)~\cite{wang2019neural} incorporates two GCNs to learn the collaborative signal of user interactions from a user-item interaction graph. 
Hyperbolic Graph Collaborative Filtering (HGCF)~\cite{sun2021hgcf} offers a compelling solution by integrating GCN with hyperbolic learning techniques to acquire user and item embeddings within the hyperbolic space. By leveraging the exponential neighborhood expansion inherent in the hyperbolic space, HGCF effectively captures higher-order relationships among users and items, enhancing the learning capabilities for downstream CF models.
GC-MC~\cite{berg2017graph} uses a GCN-based auto-encoder to learn latent features of users and items from an interaction graph and reconstructs the rating links for matrix completion. 
Later, LightGCN~\cite{he2020lightgcn} simplifies the GCN in recommendation task by only including neighborhood aggregation for calculating user and item representations, which further boosts the efficiency of subsequent GCF approaches, e.g.,~\cite{wang2020disentangled,xia2022hypergraph,sun2021hgcf,lee2021bootstrapping}.

Existing GCN-based GCF methods only capture correlation signals of user behaviors by modeling neighboring node messages.
This would result in the limited ability of GCF models to capture the true user preferences in the presence of spurious correlations.
On the contrary, we abandon the modeling of spurious correlations to pursue the intrinsic causal relations between nodes, which estimate the causal effect of a specific item on user preferences to uncover true user interests.

\subsection{Causal Modeling for Recommendation}

Recent recommendation research has largely favored causality-driven methods.
A burst of relevant papers is proposed to address critical issues in RS, such as data bias and model explainability with causal learning.
Among them, the Structural Causal Model (SCM) from Pearl et al.~\cite{pearl2009causality} has been intensively investigated.
SCM-based recommendation builds a graphical Causal Graph by extracting structural equations on causal relations between deterministic variables in recommendations.
It aims to use the Causal Graph to conduct causal reasoning for causal effect estimation.
Using the Causal Graph, most relevant approaches pursue mitigating the bad effects of different data biases, e.g., exposure bias~\cite{wang2018deconfounded,li2022causal}, popularity bias~\cite{zheng2021disentangling,zhang2021causal}. 
For instance, Wang et al.~\cite{wang2018deconfounded} mitigate exposure bias in the partially observed user-item interactions by regarding the bias as the confounder in the Causal Graph.
They propose a decounfonded model that performs Poisson factorization on substitute confounders (i.e., an exposure matrix) and partially observed user ratings. 
Zheng et al.~\cite{zheng2021disentangling} relate the user conformity issue in recommendations with popularity bias, and use a Causal Graph to guide the disentangled learning of user interest embeddings. 
Other approaches also achieve explainable recommendations.
Wang et al.~\cite{wang2022causal} define a Causal Graph that shows how users' true intents are related to item semantics, i.e., attributes.
They propose a framework that produces disentangled semantics-aware user intent embeddings, in which each model component corresponds to a specific node in the Causal Graph.
The learned embeddings are able to disentangle users' true intents towards specific item semantics, which explains which item attributes are favored by users.

\section{Conclusion}
We propose NCGCF, the first causality-aware graph representation learning framework for collaborative filtering.
Our NCGCF injects causal relations between nodes into GCN-based graph representation learning to derive satisfactory user and item representations for the CF model.
We craft a Causal Graph to describe the causality-aware graph representation learning process.
Our NCGCF quantifies each of the structural equations under the Causal Graph, with a semi-implicit generative model enabling causality-aware message passing for graph learning.
Finally, NCGCF produces causality-aware graph embeddings by modeling dependencies of structural equations, thus enabling better user preference modeling.
Extensive evaluations on four datasets demonstrate NCGCF’s ability to produce precise recommendations that interpret user preferences and uncover user behavior patterns.



\bibliographystyle{IEEEtran}
\bibliography{bibs}

\begin{thebibliography}{10}
\providecommand{\url}[1]{#1}
\csname url@samestyle\endcsname
\providecommand{\newblock}{\relax}
\providecommand{\bibinfo}[2]{#2}
\providecommand{\BIBentrySTDinterwordspacing}{\spaceskip=0pt\relax}
\providecommand{\BIBentryALTinterwordstretchfactor}{4}
\providecommand{\BIBentryALTinterwordspacing}{\spaceskip=\fontdimen2\font plus
\BIBentryALTinterwordstretchfactor\fontdimen3\font minus
  \fontdimen4\font\relax}
\providecommand{\BIBforeignlanguage}[2]{{%
\expandafter\ifx\csname l@#1\endcsname\relax
\typeout{** WARNING: IEEEtran.bst: No hyphenation pattern has been}%
\typeout{** loaded for the language `#1'. Using the pattern for}%
\typeout{** the default language instead.}%
\else
\language=\csname l@#1\endcsname
\fi
#2}}
\providecommand{\BIBdecl}{\relax}
\BIBdecl

\bibitem{schafer2007collaborative}
J.~B. Schafer, D.~Frankowski, J.~Herlocker, and S.~Sen, ``Collaborative
  filtering recommender systems,'' in \emph{The adaptive web}.\hskip 1em plus
  0.5em minus 0.4em\relax Springer, 2007, pp. 291--324.

\bibitem{xia2021graph}
F.~Xia, K.~Sun, S.~Yu, A.~Aziz, L.~Wan, S.~Pan, and H.~Liu, ``Graph learning: A
  survey,'' \emph{IEEE Transactions on Artificial Intelligence}, vol.~2, no.~2,
  pp. 109--127, 2021.

\bibitem{wang2021graph}
S.~Wang, L.~Hu, Y.~Wang, X.~He, Q.~Z. Sheng, M.~A. Orgun, L.~Cao, F.~Ricci, and
  P.~S. Yu, ``Graph learning based recommender systems: A review,'' \emph{arXiv
  preprint arXiv:2105.06339}, 2021.

\bibitem{chen2018matrix}
S.~Chen and Y.~Peng, ``Matrix factorization for recommendation with explicit
  and implicit feedback,'' \emph{Knowledge-Based Systems}, vol. 158, pp.
  109--117, 2018.

\bibitem{he2017neural}
X.~He, L.~Liao, H.~Zhang, L.~Nie, X.~Hu, and T.-S. Chua, ``Neural collaborative
  filtering,'' in \emph{Proceedings of the 26th international conference on
  world wide web}, 2017, pp. 173--182.

\bibitem{hamilton2020graph}
W.~L. Hamilton, ``Graph representation learning,'' \emph{Synthesis Lectures on
  Artifical Intelligence and Machine Learning}, vol.~14, no.~3, pp. 1--159,
  2020.

\bibitem{wang2019neural}
X.~Wang, X.~He, M.~Wang, F.~Feng, and T.-S. Chua, ``Neural graph collaborative
  filtering,'' in \emph{Proceedings of the 42nd international ACM SIGIR
  conference on Research and development in Information Retrieval}, 2019, pp.
  165--174.

\bibitem{sun2021hgcf}
J.~Sun, Z.~Cheng, S.~Zuberi, F.~P{\'e}rez, and M.~Volkovs, ``Hgcf: Hyperbolic
  graph convolution networks for collaborative filtering,'' in
  \emph{Proceedings of the Web Conference 2021}, 2021, pp. 593--601.

\bibitem{berg2017graph}
R.~v.~d. Berg, T.~N. Kipf, and M.~Welling, ``Graph convolutional matrix
  completion,'' \emph{arXiv preprint arXiv:1706.02263}, 2017.

\bibitem{he2020lightgcn}
X.~He, K.~Deng, X.~Wang, Y.~Li, Y.~Zhang, and M.~Wang, ``Lightgcn: Simplifying
  and powering graph convolution network for recommendation,'' in
  \emph{Proceedings of the 43rd International ACM SIGIR conference on research
  and development in Information Retrieval}, 2020, pp. 639--648.

\bibitem{yan2022two}
Y.~Yan, M.~Hashemi, K.~Swersky, Y.~Yang, and D.~Koutra, ``Two sides of the same
  coin: Heterophily and oversmoothing in graph convolutional neural networks,''
  in \emph{2022 IEEE International Conference on Data Mining (ICDM)}.\hskip 1em
  plus 0.5em minus 0.4em\relax IEEE, 2022, pp. 1287--1292.

\bibitem{10.1145/3485447.3512072}
\BIBentryALTinterwordspacing
X.~Wang, Q.~Li, D.~Yu, and G.~Xu, ``Off-policy learning over heterogeneous
  information for recommendation,'' ser. WWW '22.\hskip 1em plus 0.5em minus
  0.4em\relax New York, NY, USA: Association for Computing Machinery, 2022, p.
  2348–2359. [Online]. Available:
  \url{https://doi.org/10.1145/3485447.3512072}
\BIBentrySTDinterwordspacing

\bibitem{xu2023causal}
S.~Xu, Y.~Ge, Y.~Li, Z.~Fu, X.~Chen, and Y.~Zhang, ``Causal collaborative
  filtering,'' in \emph{Proceedings of the 2023 ACM SIGIR International
  Conference on Theory of Information Retrieval}, 2023, pp. 235--245.

\bibitem{zheng2021disentangling}
Y.~Zheng, C.~Gao, X.~Li, X.~He, Y.~Li, and D.~Jin, ``Disentangling user
  interest and conformity for recommendation with causal embedding,'' in
  \emph{Proceedings of the Web Conference 2021}, 2021, pp. 2980--2991.

\bibitem{wang2022causaldisen}
X.~Wang, Q.~Li, D.~Yu, P.~Cui, Z.~Wang, and G.~Xu, ``Causal disentanglement for
  semantics-aware intent learning in recommendation,'' \emph{IEEE Transactions
  on Knowledge and Data Engineering}, 2022.

\bibitem{bareinboim2022pearl}
E.~Bareinboim, J.~D. Correa, D.~Ibeling, and T.~Icard, ``On pearl’s hierarchy
  and the foundations of causal inference,'' in \emph{Probabilistic and Causal
  Inference: The Works of Judea Pearl}, 2022, pp. 507--556.

\bibitem{xia2021causal}
K.~Xia, K.-Z. Lee, Y.~Bengio, and E.~Bareinboim, ``The causal-neural
  connection: Expressiveness, learnability, and inference,'' \emph{Advances in
  Neural Information Processing Systems}, vol.~34, pp. 10\,823--10\,836, 2021.

\bibitem{pearl2000models}
J.~Pearl \emph{et~al.}, ``Models, reasoning and inference,'' \emph{Cambridge,
  UK: CambridgeUniversityPress}, vol.~19, no.~2, 2000.

\bibitem{mnih2007probabilistic}
A.~Mnih and R.~R. Salakhutdinov, ``Probabilistic matrix factorization,''
  \emph{Advances in neural information processing systems}, vol.~20, 2007.

\bibitem{agarwal2009regression}
D.~Agarwal and B.-C. Chen, ``Regression-based latent factor models,'' in
  \emph{Proceedings of the 15th ACM SIGKDD international conference on
  Knowledge discovery and data mining}, 2009, pp. 19--28.

\bibitem{wang2018deconfounded}
Y.~Wang, D.~Liang, L.~Charlin, and D.~M. Blei, ``The deconfounded recommender:
  A causal inference approach to recommendation,'' \emph{arXiv preprint
  arXiv:1808.06581}, 2018.

\bibitem{bhadani2021biases}
S.~Bhadani, ``Biases in recommendation system,'' in \emph{Proceedings of the
  15th ACM Conference on Recommender Systems}, 2021, pp. 855--859.

\bibitem{wang2020disentangled}
X.~Wang, H.~Jin, A.~Zhang, X.~He, T.~Xu, and T.-S. Chua, ``Disentangled graph
  collaborative filtering,'' in \emph{Proceedings of the 43rd international ACM
  SIGIR conference on research and development in information retrieval}, 2020,
  pp. 1001--1010.

\bibitem{li2022causal}
Q.~Li, X.~Wang, Z.~Wang, and G.~Xu, ``Be causal: De-biasing social network
  confounding in recommendation,'' \emph{ACM Transactions on Knowledge
  Discovery from Data (TKDD)}, 2022.

\bibitem{pearl2018book}
J.~Pearl and D.~Mackenzie, \emph{The book of why: the new science of cause and
  effect}.\hskip 1em plus 0.5em minus 0.4em\relax Basic books, 2018.

\bibitem{kipf2016variational}
T.~N. Kipf and M.~Welling, ``Variational graph auto-encoders,'' \emph{arXiv
  preprint arXiv:1611.07308}, 2016.

\bibitem{yin2018semi}
M.~Yin and M.~Zhou, ``Semi-implicit variational inference,'' in
  \emph{International Conference on Machine Learning}.\hskip 1em plus 0.5em
  minus 0.4em\relax PMLR, 2018, pp. 5660--5669.

\bibitem{hasanzadeh2019semi}
A.~Hasanzadeh, E.~Hajiramezanali, K.~Narayanan, N.~Duffield, M.~Zhou, and
  X.~Qian, ``Semi-implicit graph variational auto-encoders,'' \emph{Advances in
  neural information processing systems}, vol.~32, 2019.

\bibitem{8588399}
C.~Zhang, J.~Bütepage, H.~Kjellström, and S.~Mandt, ``Advances in variational
  inference,'' \emph{IEEE Transactions on Pattern Analysis and Machine
  Intelligence}, vol.~41, no.~8, pp. 2008--2026, 2019.

\bibitem{altaf2019dataset}
B.~Altaf, U.~Akujuobi, L.~Yu, and X.~Zhang, ``Dataset recommendation via
  variational graph autoencoder,'' in \emph{2019 IEEE International Conference
  on Data Mining (ICDM)}.\hskip 1em plus 0.5em minus 0.4em\relax IEEE, 2019,
  pp. 11--20.

\bibitem{xu2018representation}
K.~Xu, C.~Li, Y.~Tian, T.~Sonobe, K.-i. Kawarabayashi, and S.~Jegelka,
  ``Representation learning on graphs with jumping knowledge networks,'' in
  \emph{International Conference on Machine Learning}.\hskip 1em plus 0.5em
  minus 0.4em\relax PMLR, 2018, pp. 5453--5462.

\bibitem{koren2008factorization}
Y.~Koren, ``Factorization meets the neighborhood: a multifaceted collaborative
  filtering model,'' in \emph{Proceedings of the 14th ACM SIGKDD international
  conference on Knowledge discovery and data mining}, 2008, pp. 426--434.

\bibitem{koren2022advances}
Y.~Koren, S.~Rendle, and R.~Bell, ``Advances in collaborative filtering,''
  \emph{Recommender systems handbook}, pp. 91--142, 2022.

\bibitem{wang2022mgpolicy}
X.~Wang, Q.~Li, D.~Yu, Z.~Wang, H.~Chen, and G.~Xu, ``Mgpolicy: Meta graph
  enhanced off-policy learning for recommendations,'' in \emph{Proceedings of
  the 45th International ACM SIGIR Conference on Research and Development in
  Information Retrieval}, 2022, pp. 1369--1378.

\bibitem{xiong2021counterfactual}
K.~Xiong, W.~Ye, X.~Chen, Y.~Zhang, W.~X. Zhao, B.~Hu, Z.~Zhang, and J.~Zhou,
  ``Counterfactual review-based recommendation,'' in \emph{Proceedings of the
  30th ACM International Conference on Information \& Knowledge Management},
  2021, pp. 2231--2240.

\bibitem{zhang2020causal}
C.~Zhang, K.~Zhang, and Y.~Li, ``A causal view on robustness of neural
  networks,'' \emph{Advances in Neural Information Processing Systems},
  vol.~33, pp. 289--301, 2020.

\bibitem{liang2018variational}
D.~Liang, R.~G. Krishnan, M.~D. Hoffman, and T.~Jebara, ``Variational
  autoencoders for collaborative filtering,'' in \emph{Proceedings of the 2018
  world wide web conference}, 2018, pp. 689--698.

\bibitem{van2001art}
D.~A. Van~Dyk and X.-L. Meng, ``The art of data augmentation,'' \emph{Journal
  of Computational and Graphical Statistics}, vol.~10, no.~1, pp. 1--50, 2001.

\bibitem{barkan2016item2vec}
O.~Barkan and N.~Koenigstein, ``Item2vec: neural item embedding for
  collaborative filtering,'' in \emph{2016 IEEE 26th International Workshop on
  Machine Learning for Signal Processing (MLSP)}.\hskip 1em plus 0.5em minus
  0.4em\relax IEEE, 2016, pp. 1--6.

\bibitem{he2016ups}
R.~He and J.~McAuley, ``Ups and downs: Modeling the visual evolution of fashion
  trends with one-class collaborative filtering,'' in \emph{proceedings of the
  25th international conference on world wide web}, 2016, pp. 507--517.

\bibitem{haque2018sentiment}
T.~U. Haque, N.~N. Saber, and F.~M. Shah, ``Sentiment analysis on large scale
  amazon product reviews,'' in \emph{2018 IEEE international conference on
  innovative research and development (ICIRD)}.\hskip 1em plus 0.5em minus
  0.4em\relax IEEE, 2018, pp. 1--6.

\bibitem{tang-etal12a}
J.~Tang, H.~Gao, and H.~Liu, ``m{T}rust: {D}iscerning multi-faceted trust in a
  connected world,'' in \emph{Proceedings of the fifth ACM international
  conference on Web search and data mining}.\hskip 1em plus 0.5em minus
  0.4em\relax ACM, 2012, pp. 93--102.

\bibitem{rendle2012bpr}
S.~Rendle, C.~Freudenthaler, Z.~Gantner, and L.~Schmidt-Thieme, ``Bpr: Bayesian
  personalized ranking from implicit feedback,'' \emph{arXiv preprint
  arXiv:1205.2618}, 2012.

\bibitem{zhang2021causally}
J.~Zhang, X.~Chen, and W.~X. Zhao, ``Causally attentive collaborative
  filtering,'' in \emph{Proceedings of the 30th ACM International Conference on
  Information \& Knowledge Management}, 2021, pp. 3622--3626.

\bibitem{doi:https://doi.org/10.1002/9780471462422.eoct979}
\BIBentryALTinterwordspacing
R.~F. Woolson, \emph{Wilcoxon Signed-Rank Test}.\hskip 1em plus 0.5em minus
  0.4em\relax John Wiley \& Sons, Ltd, 2008, pp. 1--3. [Online]. Available:
  \url{https://onlinelibrary.wiley.com/doi/abs/10.1002/9780471462422.eoct979}
\BIBentrySTDinterwordspacing

\bibitem{kipf2016semi}
T.~N. Kipf and M.~Welling, ``Semi-supervised classification with graph
  convolutional networks,'' \emph{arXiv preprint arXiv:1609.02907}, 2016.

\bibitem{hamilton2017inductive}
W.~Hamilton, Z.~Ying, and J.~Leskovec, ``Inductive representation learning on
  large graphs,'' \emph{Advances in neural information processing systems},
  vol.~30, 2017.

\bibitem{ying2018graph}
R.~Ying, R.~He, K.~Chen, P.~Eksombatchai, W.~L. Hamilton, and J.~Leskovec,
  ``Graph convolutional neural networks for web-scale recommender systems,'' in
  \emph{Proceedings of the 24th ACM SIGKDD international conference on
  knowledge discovery \& data mining}, 2018, pp. 974--983.

\bibitem{xia2022hypergraph}
L.~Xia, C.~Huang, Y.~Xu, J.~Zhao, D.~Yin, and J.~Huang, ``Hypergraph
  contrastive collaborative filtering,'' in \emph{Proceedings of the 45th
  International ACM SIGIR conference on research and development in information
  retrieval}, 2022, pp. 70--79.

\bibitem{lee2021bootstrapping}
D.~Lee, S.~Kang, H.~Ju, C.~Park, and H.~Yu, ``Bootstrapping user and item
  representations for one-class collaborative filtering,'' in \emph{Proceedings
  of the 44th International ACM SIGIR Conference on Research and Development in
  Information Retrieval}, 2021, pp. 317--326.

\bibitem{pearl2009causality}
J.~Pearl, \emph{Causality}.\hskip 1em plus 0.5em minus 0.4em\relax Cambridge
  university press, 2009.

\bibitem{zhang2021causal}
Y.~Zhang, F.~Feng, X.~He, T.~Wei, C.~Song, G.~Ling, and Y.~Zhang, ``Causal
  intervention for leveraging popularity bias in recommendation,'' \emph{arXiv
  preprint arXiv:2105.06067}, 2021.

\bibitem{wang2022causal}
W.~Wang, X.~Lin, F.~Feng, X.~He, M.~Lin, and T.-S. Chua, ``Causal
  representation learning for out-of-distribution recommendation,'' in
  \emph{Proceedings of the ACM Web Conference 2022}, 2022, pp. 3562--3571.

\end{thebibliography}

\end{document}